\documentclass[aip,apl,reprint,twocolumn]{revtex4-1}
\usepackage{graphicx} 
\usepackage{bm} 
\usepackage{dcolumn}
\usepackage{amssymb,amsmath}
\bibstyle{apsrev}
\usepackage{hyperref}
\usepackage{color}
\usepackage[normalem]{ulem}

\begin{document}
\title{Frequency linewidth and decay length of spin waves in curved magnetic membranes}
\author{J. A. Ot\'alora}
\email[Corresponding author: ]{j.otalora@ifw-dresden.de}
\affiliation{Institute of Metallic Materials at the Leibniz Institute for Solid State and Materials Research, IFW, 01069 Dresden, Germany}
\author{A. K\'akay}
\affiliation{Helmholtz-Zentrum Dresden - Rossendorf, Institute of Ion Beam Physics and Materials Research,
Bautzner Landstr. 400, 01328 Dresden, Germany}
\author{J.  Lindner}
\affiliation{Helmholtz-Zentrum Dresden - Rossendorf, Institute of Ion Beam Physics and Materials Research,
Bautzner Landstr. 400, 01328 Dresden, Germany}
\author{H. Schultheiss}
\affiliation{Helmholtz-Zentrum Dresden - Rossendorf, Institute of Ion Beam Physics and Materials Research,
Bautzner Landstr. 400, 01328 Dresden, Germany}
\author{A. Thomas}
\affiliation{Institute of Metallic Materials at the Leibniz Institute for Solid State and Materials Research, IFW, 01069 Dresden, Germany}
\author{J. Fassbender}
\affiliation{Helmholtz-Zentrum Dresden - Rossendorf, Institute of Ion Beam Physics and Materials Research,
Bautzner Landstr. 400, 01328 Dresden, Germany}
\author{K. Nielsch}
\affiliation{Institute of Metallic Materials at the Leibniz Institute for Solid State and Materials Research, IFW, 01069 Dresden, Germany}
\affiliation{Technische Universit\"aˆt Dresden, Institute of Materials Science, 01062 Dresden, Germany}
\date{\today}
\begin{abstract}
%The toolbox of mechanisms and physical effects for controlling spin-waves is being expanded 
The curvature of a magnetic membrane was presented as a means of inducing nonreciprocities in the spin-wave (SW) dispersion relation (see [Ot\'alora \textit{et al. Phys. Rev. Lett.}, 2016 \textbf{117}, 227203] and [Ot\'alora \textit{et al. Phys. Rev. B.}, 2017 \textbf{95}, 184415]), thereby expanding the toolbox for controlling SWs. In this paper, we further complement this toolbox by analytically showing that the membrane curvature is also manifested in the absorption of SWs, leading to a difference in the frequency linewidth (or lifetime) of counterpropagating magnons. Herein, we studied the nanotubular case, predicting changes of approximately greater than 10\% and up to 20\% in the frequency linewidth of counterpropagating SWs for a wide range of nanotube radii ranging from $30$ nm to $260$ nm and with a thickness of $10$ nm. These percentages are comparable to those that can be extracted from experiments on heavy metal/magnetic metal sandwiches, wherein linewidth asymmetry results from an interfacial Dzyaloshinskii-Moriya interaction (DMI). We also show that the interplay between the frequency linewidth and group velocity leads to asymmetries in the SW decay length, presenting changes between 10\% and 22\% for counterpropagating SWs in the frequency range of 2 - 10 GHz. For the case of the SW dispersion relation, the predicted effects are identified as the classical dipole-dipole interaction, and the analytical expression of the frequency linewidth has the same mathematical form as in thin films with the DMI. Furthermore, we present limiting cases of a tubular geometry with negligible curvature such that our analytical model converges to the case of a planar thin film known from the literature. Our findings represent a step forward toward the realization of three-dimensional curvilinear magnonic devices.

\end{abstract}
\pagestyle{plain} 
\maketitle
 
\section{Introduction}

In Magnonics, spin waves (SWs) or magnons -- the eigenoscillations of an electron spin system\cite{BlochZP30} -- are considered to be the elemental information carriers. Since magnons can propagate up to terahertz (THz) frequencies with nanometric wavelengths and over macroscopic distances without electron charges being displaced, applications based on magnons could be exempt from Joule heating, thus paving the way toward applications with unprecedentedly low power consumption (energy-friendly environmental devices), reconfigurable functionality, faster operation and further miniaturization. Remarkable progress has been achieved both theoretically and experimentally, leading to advances in the circuitry of magnonic chips, for instance, in reconfigurable waveguides for on-demand control of SWs\cite{GRUNDNPHYS15,GrundlerNatNan16}, and consequently in prototype building blocks of SW-based logic elements.\cite{Chumak14,Vogt_APL12,Chumak14MT,Vogt14,Lenk11}

%+++++++++ Figure1: Illustration +++++++++++
\begin{figure}[h!]
\begin{center}
\includegraphics[scale=0.6,angle=0]{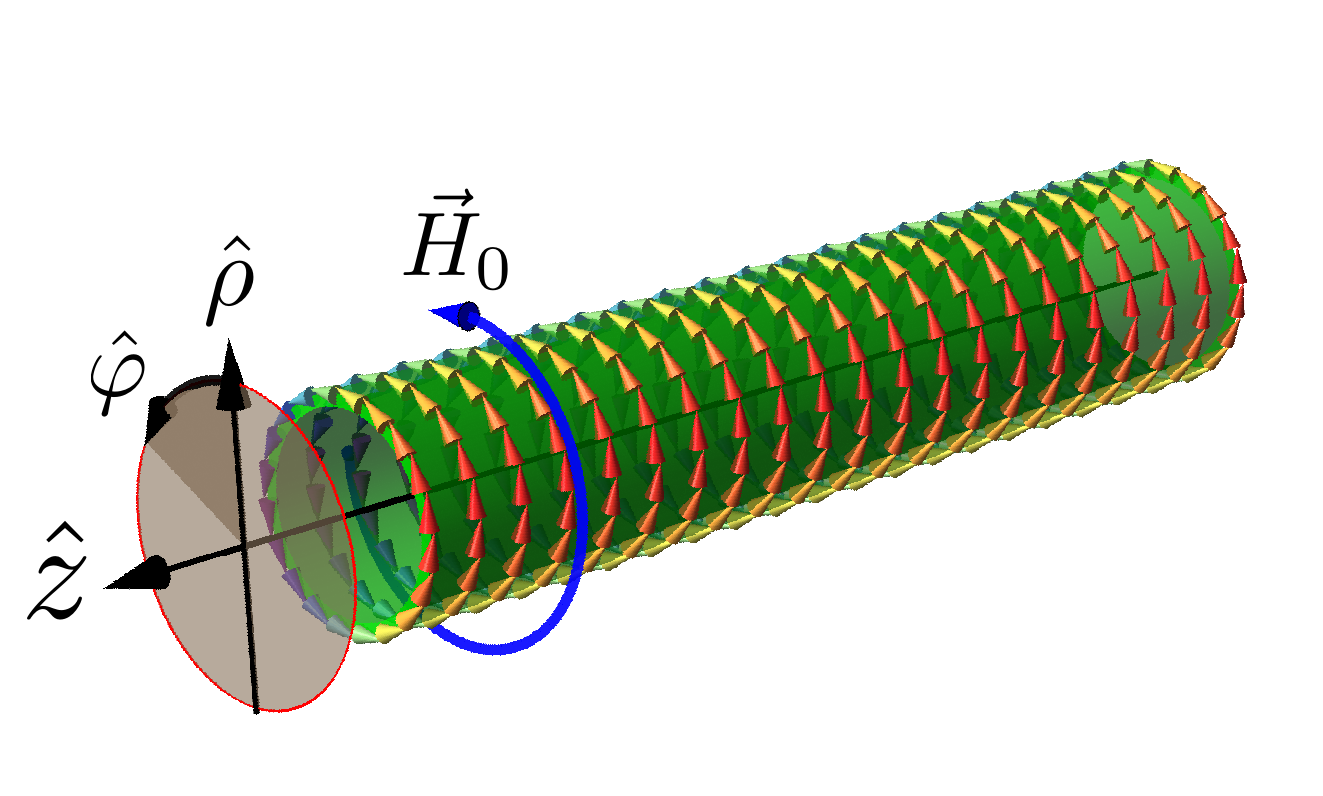}
\end{center}
\caption{(Color online). Schematic representation of a nanotube in a circular (vortex) magnetic configuration $\vec M = M_s\hat\varphi$. For nanotubes with radii smaller than a critical radius, a circular (azimuthal) magnetic field $\vec H_0=H_0 \hat\varphi$ is required to stabilize this configuration, while the exchange interaction prefers the homogeneous magnetization state along the long axis. The critical magnetic field related to the critical radius can be analytically calculated for a nanotube with specific geometrical and material parameters, as given by ~\textcite{OtaloraJAP15}. However, for radii above $100$ nm, for most of the metallic magnetic materials, the vortex state is the preferred equilibrium configuration.
}
\label{Illustration}
\end{figure}
%++++++++++++++++++++++++++++++++++++

From a technological perspective, a desirable feature of magnons is the nonreciprocal (or asymmetric) dispersion relation. For instance, in SW-based logic devices, the asymmetric dispersion relation provides the condition for the unidirectional propagation of SW packages (avoiding the formation of standing SWs), which is fundamental for enhancing the input/output information transfer efficiency encoded in the  phase, frequency and amplitude of SWs in the magnonic device. An asymmetric dispersion relation generally refers to the broken parity symmetry of the eigenfrequency $f$ regarding the wavevector $\vec k$, i.e., $f[\vec k]\neq f[-\vec k]$, meaning that counterpropagating SWs with the same wavevector have different oscillation frequencies. Furthermore, a nonreciprocal dispersion relation also means an asymmetric group velocity $(\vec v_g\equiv 2\pi\nabla_{\vec k}f)$. This is generally accompanied by asymmetries in lifetimes ($\tau\equiv1/\Gamma[\vec k]$) and decay lengths ($l_{D}\equiv v_g \tau$), which are also desired properties in the pursuit of enhancing performance in SW-based communications and logic device applications.\cite{ChumakNCOM14,ChumakNP15}

Substantial efforts to enhance nonreciprocities on SWs are currently taking place, mostly in two-dimensional (2D) thin films. For instance, efforts have focused on optimizing the ferromagnetic/heavy metal interface to strengthen the Dzyaloshinskii-Moriya interaction (DMI)\cite{DzyaloshinskyJPCS58,MoriyaPR60} and therefore optimize the asymmetries in frequency and linewidth.\cite{zakeri_asymmetric_2010,CorteOrtunoJPCM13,MaRSCA14,DiPRL15,IguchiPRB15} Recently, a nonreciprocal dispersion relation was predicted by turning an ordinary 2D magnetic thin film into a curvilinear three-dimensional (3D) shape, particularly into helicoids and nanotubes. It has been predicted that the topological curvature of the 3D film induces asymmetries in the SW dispersion relation, which arises from exchange interactions in helicoids and dipole-dipole interactions in nanotubes.\cite{hertel_curvature-induced_2013,GaidideiArxiv17,ShekaPRB17,OtaloraPRL16,OtaloraPRB17} It would be expected that the membrane curvature can also induce asymmetries in the frequency linewidth and decay length of SWs; nevertheless, these features have not yet been investigated, and this paper aims to fill this gap. 

Accordingly, an analytical derivation of the Polder tensor\cite{PolderMJS49} (also known as the dynamic susceptibility tensor) of the asymmetries in the frequency linewidth and decay length and the analysis of the role of curvature are presented here only for the nanotubular geometry. As reported in previous works, the mean curvature of the nanotube breaks the chiral degeneracy of the dynamic dipolar field created by the spatial distribution of the counterpropagating SWs\cite{OtaloraPRL16,OtaloraPRB17} and results in the asymmetric dispersion relation. We extend our calculation to show that, as expected, curvature-induced asymmetries in frequency linewidth and decay length also occur as a consequence of such a broken chiral degeneracy.

With the aim of clearly presenting our results, the remainder of this paper is organized as follows. The analytical model is introduced in Sec.~\ref{Analytical}, wherein explicit expressions for the Polder tensor, absorption, frequency linewidth, decay length and its asymmetries are obtained. These quantities are presented in  Sec.~\ref{Results} and discussed in Sec.~\ref{Discussions}. Furthermore, for completeness, a few limiting cases showing a direct resemblance with SWs in 2D thin films are shown and discussed in Sec.~\ref{LimitCases}. Finally, conclusions are presented in Sec.~\ref{Conclusions}

\section{ANALYTICAL MODEL}
\label{Analytical}
The nonreciprocal SW dispersion induced by curvature in cylindrical magnonic waveguides has been reported in our previous papers \cite{OtaloraPRB17,OtaloraPRL16}, wherein an analytical formula for the dispersion relation of SWs was presented for the case of a nanotube in an equilibrium circular (vortex) magnetic configuration. Fig. \ref{Illustration} illustrates the circular equilibrium state and the cylindrical coordinate system defined by the radial $\hat \rho$, azimuthal $\hat \varphi$ and longitudinal $\hat z$ unitary vectors. Using these coordinates, the vortex configuration is described by $\vec \Omega_0=\vec M/M_s=\hat \varphi$, the unitary magnetization field vector, and $M_s$, the saturation magnetization. This equilibrium magnetization and the cylindrical coordinate system will be assumed throughout the manuscript.  In the following, the methods applied in our previous papers will be extended for analytically calculating the dynamic susceptibility ($\aleph$) and frequency linewidth ($\Gamma$) in the presence of damping. In both quantities, the curvature-induced nonreciprocity is present. The SW properties are studied by linearizing the Landau-Lifshitz Gilbert (LLG) equation of motion,

\begin{equation}
\label{LLG}
\begin{aligned}
\dot{\vec \Omega }=-\gamma _0\  \vec\Omega \times\big( \vec H_{\text{eff}}+\vec h_{\text{rf}}\big) -\alpha _G \left(\dot{\vec\Omega }\times \vec\Omega \right),
\end{aligned}
\end{equation} 

\noindent where $\gamma_0=g\mu_B/\hbar$ is the gyromagnetic ratio, $g$ is the electron's Land\'e factor, $\mu_B$ is the Bohr magneton and $\hbar$ is the Planck constant. The effective field $ \vec H_{\text{eff}}=\vec H_{\text{ex}}+ \vec H_{\text{d}}+ \vec H_{\text{0}}$ is given by the contributions of (i) the exchange field $\vec H_{\text{ex}}=D \nabla^2\vec\Omega$, where $D=l_{\text{ex}}^2 M_s$, the exchange length $l_{\text{ex}}=\sqrt{A/K_d}$, the stiffness constant $A$, and $K_d=(1/2)\mu_0M_s^2$; (ii) the dipolar field $\vec H_d=-\nabla\Phi$ arising from the magnetostatic potential $\Phi$; and (iii) the circular (or azimuthal) $\vec{ H}_0=H_0\hat\varphi$ applied external magnetic field.
The first term on the right side in Eq. \ref{LLG}, known as the Landau-Lifshitz torque, describes the precession of the magnetization in the effective field. SWs are excited with a radio-frequency magnetic field $\vec h_{\text{rf}}$. The second term on the right side in Eq. \ref{LLG} is known as the Gilbert torque, where $\alpha_G$ is the Gilbert damping constant accounting for the relaxation mechanism that dissipates energy. 

The LLG equation is linearized in terms of small perturbations of the equilibrium magnetization. Accordingly, a sightly different magnetization state $\vec \Omega\approx\vec \Omega_0+\vec m$ is considered, where the radio-frequency field amplitude is assumed to be small ($\|\vec h_{\text{rf}}\|/M_s\ll1$) and $\vec m$ is the dynamic perturbation ($\|\vec m\|\ll1$) perpendicular to the equilibrium state ($\vec \Omega_0\cdot\vec m=0$). The small magnetic deviation can be expressed as $\vec m=m_\rho\hat\rho+m_z\hat z$, and by following procedures similar to those in J. Ot\'alora et al. \cite{OtaloraPRB17,OtaloraPRL16}, the dynamic equation is calculated as:

\begin{equation}
\label{dynamicEq}
\begin{aligned}
\frac{\dot{m}_{\rho }}{\omega _{M_s}}=-l_{ex}^2\left(\frac{1}{b^2}+\nabla^2\right)m_z&+\frac{H_0}{M_s} m_z+\frac{\left\langle h_d\right\rangle_z}{M_s}\\
&+\frac{\left\langle h_{\text{rf}}\right\rangle_z}{M_s}+\alpha_G\dot{m}_z, \\
\frac{\dot{m}_z}{\omega_{M_s}}=l_{ex}^2\nabla^2 m_\rho - \frac{H_0}{M_s}  m_\rho+\frac{\left\langle h_d\right\rangle_\rho}{M_s}&+\frac{\left\langle h_{\text{rf}}\right\rangle_{\rho}}{M_s}-\alpha_G\dot{m}_\rho,
\end{aligned}
\end{equation}
 
 \noindent where $\omega_{M_s}\equiv\gamma_0\mu_0 M_s$, $b^{-2}=2\pi \log(R/r)/S$, $S=\pi (R^2-r^2)$ is the nanotube cross-section area and  $\left\langle h_d\right\rangle_x=(2\pi/S)\int_r^R\rho d\rho (h_d)_x$ with $x=\rho,z$ is the cross-section average of the dynamic dipolar field components. Similarly, the radio-frequency field components are also averaged as $\left\langle h_{\text{rf}}\right\rangle_x=(2\pi/S)\int_r^R\rho d\rho (h_{\text{rf}})_x$ with $x=\rho,z$.  These averages are valid for nanotubes with small thicknesses for which the dynamical magnetization components  $\vec m_\rho$ and $\vec m_z$ are assumed to be radially invariant. Accordingly, the dynamical magnetization components in the Fourier space are written as 
 
 \begin{equation}
\label{Fourierm}
\begin{aligned}
m_{\rho }[\varphi ,z,t]&=\sum _{n=-\infty }^{\infty } \int _{-\infty }^{\infty }dk_{\text z}\ e^{i k_{\text z} z} e^{i n \varphi }e^{-i\omega t} \ \mathcal{R}_{k_z}^n ,\\
m_z[\varphi ,z,t]&=\sum _{n=-\infty }^{\infty } \int _{-\infty }^{\infty }dk_{\text z} \ e^{i k_{\text z} z} e^{i n \varphi } e^{-i\omega t} \ \mathcal{Z}_{k_z}^n \\
\end{aligned}
\end{equation}
 
 \noindent where $\mathcal{R}_{k_z}^n$ and  $\mathcal{Z}_{k_z}^n$ are the amplitude projections along the $\hat \rho$ and $\hat z$ directions, respectively. The eigen-excitations or magnons are characterized by the longitudinal wavevector $k_{\text z}$ along $\hat z$ and by the azimuthal wavenumber (or mode) $n$ along $\hat\varphi$. The averaged dipolar field can also be expanded as $\left\langle h_d\right\rangle_x=\sum _{n=-\infty }^{\infty } \int _{-\infty }^{\infty }dk_{\text z} \ e^{i k_{\text z} z} e^{i n \varphi }e^{-i\omega t} \ \left\langle h_d\right\rangle_x^{n,k_{\text z}}$ with $x=\rho, z$, where $\left\langle h_d\right\rangle_x^{n,k_{\text z}}$ are the dynamic dipolar field components in the Fourier space, and their explicit form is presented in our previous work (see equation 5 in \cite{OtaloraPRB17}).  Finally, the dynamic equation \ref{dynamicEq} for the excitation $\vec m$ in the Fourier space can be written in the following form:

\begin{equation}
\left(
\begin{array}{c}
 \mathcal{R}_{k_z}^n \\
 \mathcal{Z}_{k_z}^n \\
\end{array}
\right)
=\aleph_{k_z}^n
\left(
\begin{array}{c}
\left\langle h_{\text{rf}}\right\rangle_{\rho}^{n,k_z} \\
\left\langle h_{\text{rf}}\right\rangle_{z}^{n,k_z} \\
\end{array}
\right)
\end{equation}

\noindent where $\aleph_{k_{\text z}}^{n}$ is the dynamic susceptibility (also known as the Polder tensor\cite{PolderMJS49}), which is expressed as 

\begin{equation}
\label{DynSuscept}
\aleph_{k_z}^n=
\left(
\begin{array}{cc}
\chi_{\rho\rho}^{n,k_{\text{z}}} & \chi_{\rho z}^{n,k_{\text{z}}}  \\
- \chi_{\rho z}^{n,k_{\text{z}}}  & \chi_{zz}^{n,k_{\text{z}}} 
\end{array}
\right)
=
\left(
\begin{array}{cc}
\frac{i \mathbb{B}}{\mathbb{A}^2+\mathbb{B} \mathbb{C}} & \frac{i \mathbb{A}}{\mathbb{A}^2+\mathbb{B} \mathbb{C}}  \\
- \frac{i \mathbb{A}}{\mathbb{A}^2+\mathbb{B} \mathbb{C}} &\frac{i \mathbb{C}}{\mathbb{A}^2+\mathbb{B} \mathbb{C}}
\end{array}
\right)
\end{equation}

\noindent where $ \mathbb{A}\equiv\omega-\gamma M_s\mathcal{A}_{k_{\text z}}^n$, $\mathbb{B}\equiv-\alpha _G\omega+i\omega_{M_s}\mathcal{B}_{k_{\text z}}^n$, $\mathbb{C}=-  \alpha _G\omega+i \omega_{M_s}\mathcal{C}_{k_{\text z}}^n$ with $\mathcal{A}_{k_{\text z}}^n$, $\mathcal{B}_{k_{\text z}}^n$, and $\mathcal{C}_{k_{\text z}}^n$ given as 

\begin{equation}
\label{StiffnessFieldTube}
\begin{aligned}
\mathcal{A}_{k_z}^n=& \mathcal{K}_{k_z}^n,\\
\mathcal{B}_{k_z}^n=& l_{\text{ex}}^2 k_z^2+(n^2-1)h_u+h_0+\mathcal{L}_{k_z}^n, \\
\mathcal{C}_{k_z}^n=&l_{\text{ex}}^2 k_z^2+n^2h_u+h_0+\mathcal{J}_{k_z}^n, 
\end{aligned}
\end{equation}

\noindent where the term $h_u=H_u/M_s=l_{ex}^2/b^2\geq 0$ is the exchange field arising from the vortex state of the magnetization. This field can be viewed as a shape anisotropy with the large nanotube axis $\hat z$ as the easy axis. The terms denoted by $\mathcal{J}_{k_{\text z}}^n$, $\mathcal{K}_{k_{\text z}}^n$ and $\mathcal{L}_{k_{\text z}}^n$ are hypergeometrical functions that depend on the nanotube radius $R$ and thickness $d$, and their explicit formulations (the detailed derivation is presented in \cite{OtaloraPRB17}) are

\begin{equation}
\label{HyperfuncDef}
\begin{aligned}
\mathcal{J}_{k_z}^n&=\frac{\pi }{S}\int _0^{\infty }dq\frac{q^3}{2 \left(q^2+k_z^2\right)}\left(\Gamma _n[q]\right)^2\\
\mathcal{K}_{k_z}^n&=\frac{\pi }{S}\int _0^{\infty }dq\frac{q^2 k_z}{q^2+k_z^2}\Gamma _n(q)\Lambda_n[q]\\
\mathcal{L}_{k_z}^n&=\frac{\pi }{S}\int _0^{\infty }dq\frac{2 q k_z^2}{q^2+k_z^2}\left(\Lambda_n[q]\right)^2\\\end{aligned}
\end{equation}
\noindent with $\Lambda_n[q]=\int _r^Rd\rho \ \rho J_n[q \rho ]$, $\Gamma _n[q]=\Lambda_{n-1}[q]-\Lambda_{n+1}[q]$, and $J_n[x]$ the first kind Bessel functions of order $n$.

Note that the term $\mathcal{A}_{k_{\text z}}^n$ emerges only from the dynamic dipolar fields induced by the nanotubes' curvature and vanishes for planar films. As shown in our previous works  \cite{OtaloraPRB17,OtaloraPRL16}, this term is the origin for the asymmetric dispersion relation and will play the same role in frequency linewidth. Moreover, note that $\mathcal{A}_{k_{\text z}}^n$, $\mathcal{B}_{k_{\text z}}^n$ and $\mathcal{C}_{k_{\text z}}^{n}$ are the normalized stiffness fields consisting of dynamic and static components of  the exchange, magnetostatic and external fields. These features will be discussed later in section \ref{Discussions}, \textit{Discussions}.

The SW absorption (or dynamical susceptibility) can be expressed in the following Lorentzian form by using the Polder tensor defined in Eq. \ref{DynSuscept} as $\mathcal{S}_{\rho\rho}^{n,k_{\text z}}(\omega)=\text{Im}(\mathcal{X}_{\rho \rho }^{n,k_z})$

\begin{equation}
\label{absorption}
\mathcal{S}_{\rho\rho}^{n,k_{\text z}}[\omega]=\frac{\mathcal{W}_{k_{\text z}}^n[\omega,\alpha_G]}{\left(\omega{}^2-\left(\omega^{n}_{k_{\text{z}}}\right){}^2\right)^2+\left(\frac{1}{2}\Delta _{k_{\text z}}^{n}[\omega]\right)^2}
\end{equation}

\noindent where the eigenfrequency $\omega_{k_{\text z}}^{n}$ and the terms $\Delta_{k_{\text z}}^{n }(\omega)$ and $\mathcal{W}_{k_{\text z}}^{n}(\omega,\alpha_G)$ are given as

\begin{equation}
\label{dipersionrelation}
\omega^{n }_{k_{\text{z}}}=\omega_{M_s}\left(\mathcal{A}_{k_{\text z}}^n+ \sqrt{\mathcal{C}_{k_{\text z}}^n\mathcal{B}_{\text z}^n}\right),
\end{equation}

\begin{equation}
\label{linewidth}
\begin{aligned}
\Delta_{k_z}^{n }[\omega]=2 \omega_{M_s} \alpha _G \omega &\left(\mathcal{C}^{n}_{k_{\text z}}+\mathcal{B}^{n}_{k_{\text z}}\right)\left(\frac{\omega+\omega^{n }_{k_{\text{z}}}}{\omega-\bar \omega^{n }_{k_{\text{z}}}}\right)
\end{aligned}
\end{equation}

\noindent and

\begin{equation}
\begin{aligned}
\mathcal{W}_{k_{\text z}}^{n}=-\frac{\left(\omega+\omega^{n}_{k_{\text{z}}}\right)}{\left(\omega-\bar \omega^{n }_{k_{\text{z}}}\right)}\bigg(\alpha_G\omega_{M_s}\omega&\left(\omega^2-\left(\omega^{n}_{k_{\text{z}}}\right)^2\right)\\
&+\frac{(\omega_{M_s})^2}{2}\Delta _{k_z}^{n}\mathcal{B}^{n}_{k_{\text z}}\bigg),
\end{aligned}
\end{equation}

\noindent with $\bar \omega^{n }_{k_{\text{z}}}=\omega_{M_s}\left(\mathcal{A}_{k_{\text z}}^n- \sqrt{\mathcal{C}_{k_{\text z}}^n\mathcal{B}_{\text z}^n}\right)$. For planar thin films, the term $\mathcal{A}_{k_{\text z}}^n$ vanishes, and equation~\ref{dipersionrelation} simplifies to the dispersion relation that is well known from the literature.\cite{OtaloraPRL16,OtaloraPRB17} The analytical expression for the group velocity can be obtained from the $k_z$-derivative of the eigenfrequency, $v_g[n,k_z]=\partial \omega^n_{k_z}/\partial k_z$, and the frequency linewidth $\Gamma$ can be obtained from the absorption equation \ref{absorption}. Note that the Lorentzian susceptibility should be inversely proportional to the linewidth as a quadratic term, i.e., $\mathcal{S}_{\rho\rho}^{n,k_{\text z}}[\omega]\propto\left(\left(\omega-\omega^{n}_{k_{\text{z}}}\right)^2-(\Gamma/2)^2\right)^{-1}$. Hence, from equations \ref{absorption} and \ref{linewidth}, $\big($when $\Delta_{k_z}^{n }[\omega]/\left(\omega+\omega^{n}_{k_{\text{z}}}\right)$ is evaluated for the eigenfrequencies $\omega=\omega^{n }_{k_{\text{z}}}$$\big)$, it is straightforward to deduce the equation of the frequency linewidth as follows:

\begin{equation}
\label{freqLinewidth}
\Gamma[n,k_z]=\omega_{M_s} \alpha _G \left(\mathcal{C}^{n}_{k_{\text z}}+\mathcal{B}^{n}_{k_{\text z}}\right)\left(1+\frac{\mathcal{A}_{k_{\text z}}^n}{\sqrt{\mathcal{B}_{k_{\text z}}^n\mathcal{C}_{k_{\text z}}^n}}\right)
\end{equation}

This expression reduces to that of the frequency linewidth of 2D thin films when the nanotube radius goes to infinity ($\mathcal{A}_{k_{\text z}}^n=0$). In this case, the linewidth directly depends on the stiffness fields and the Gilbert damping parameter as well known from the literature~\cite{HurbenJAP98,AriasPRB99,LanderosPRB08,GallardoNJP14,DiPRL15}. For nanotubes with finite radii, equation~\ref{freqLinewidth} takes a similar mathematical form as that derived for planar 2D heavy metal/magnetic metal sandwiches with interfacial DMI (see equation 8 in \cite{DiPRL15}). The asymmetries in $\Gamma$ are introduced by the broken mirror symmetry of $\mathcal{A}_{k_{\text z}}^n$, i.e., $\mathcal{A}_{k_{\text z}}^n=-\mathcal{A}_{-k_{\text z}}^n$, which is the same term giving the asymmetries of the dispersion relation in nanotubes (see equation \ref{dipersionrelation} and equations 6 and 10 in \cite{OtaloraPRB17}). This term arises from the broken mirror symmetry of the dynamic dipolar field, which can be understood from the role of the mean curvature of the nanotube ($1/\bar\rho=2/(R+r)$) in breaking the mirror symmetry of the dynamic magnetic charges that create the dynamic dipolar field. Here, $R$ and $r$ are the outer and inner radii of a nanotube, respectively.
The asymmetry on the volume charges was previously used to explain the curvature-induced asymmetric dispersion relation in nanotubes \cite{OtaloraPRL16,OtaloraPRB17} and will be used again in the section \ref{Discussions}, \textit{Discussions}, to explain  the asymmetries in the frequency linewidth in more detail. 

The frequency linewidth asymmetry can be defined as $\Delta\Gamma[n,k_z]\equiv |\Gamma[n,k_z]-\Gamma[n,-k_z]|$ and has the form of

\begin{equation}
\label{freqLinewidthAsymmet}
\Delta\Gamma[n,k_z]=2\omega_{M_s} \alpha _G |\mathcal{C}^{n}_{k_{\text z}}+\mathcal{B}^{n}_{k_{\text z}}|\frac{|\mathcal{A}_{k_{\text z}}^n|}{\sqrt{\mathcal{B}_{k_{\text z}}^n\mathcal{C}_{k_{\text z}}^n}}
\end{equation}

\noindent Note that the frequency linewidth asymmetry is zero ($\Delta\Gamma[n,k_z]=0$) for $k_z=0$ as a consequence of the odd parity of $\mathcal{A}_{k_{\text z}}^n$. The results of the frequency linewidth and its asymmetry will be presented in the next section \ref{Results}, \textit{Results}.

\section{Results}
\label{Results}
In the following, the results of the analytical model are presented assuming a permalloy nanotube with outer (inner) radius $R=80$ nm ($r=70$ nm), saturation magnetization $\mu_0 M_s=1$ T, exchange stiffness constant $A=13$ pJ/m and exchange length $l_{\text{ex}}\approx5.8$ nm. The preferred equilibrium state for these material parameters and geometrical dimension in the absence of external fields is the saturated state along the long axis of the nanotube. 
The critical field to stabilize the vortex magnetic state is calculated to be $\mu_0 H_{\text{crit}}=\mu_0 H_u\approx 5.9$ mT.\cite{OtaloraJAP15} Therefore, a field of  $\mu_0 H_0 = 6$ mT is applied to set the circular magnetization state.

%+++++++++ Figure2: DispersionRelation +++++++++++
\begin{figure}[th!]
\begin{center}
\includegraphics[width=\linewidth]{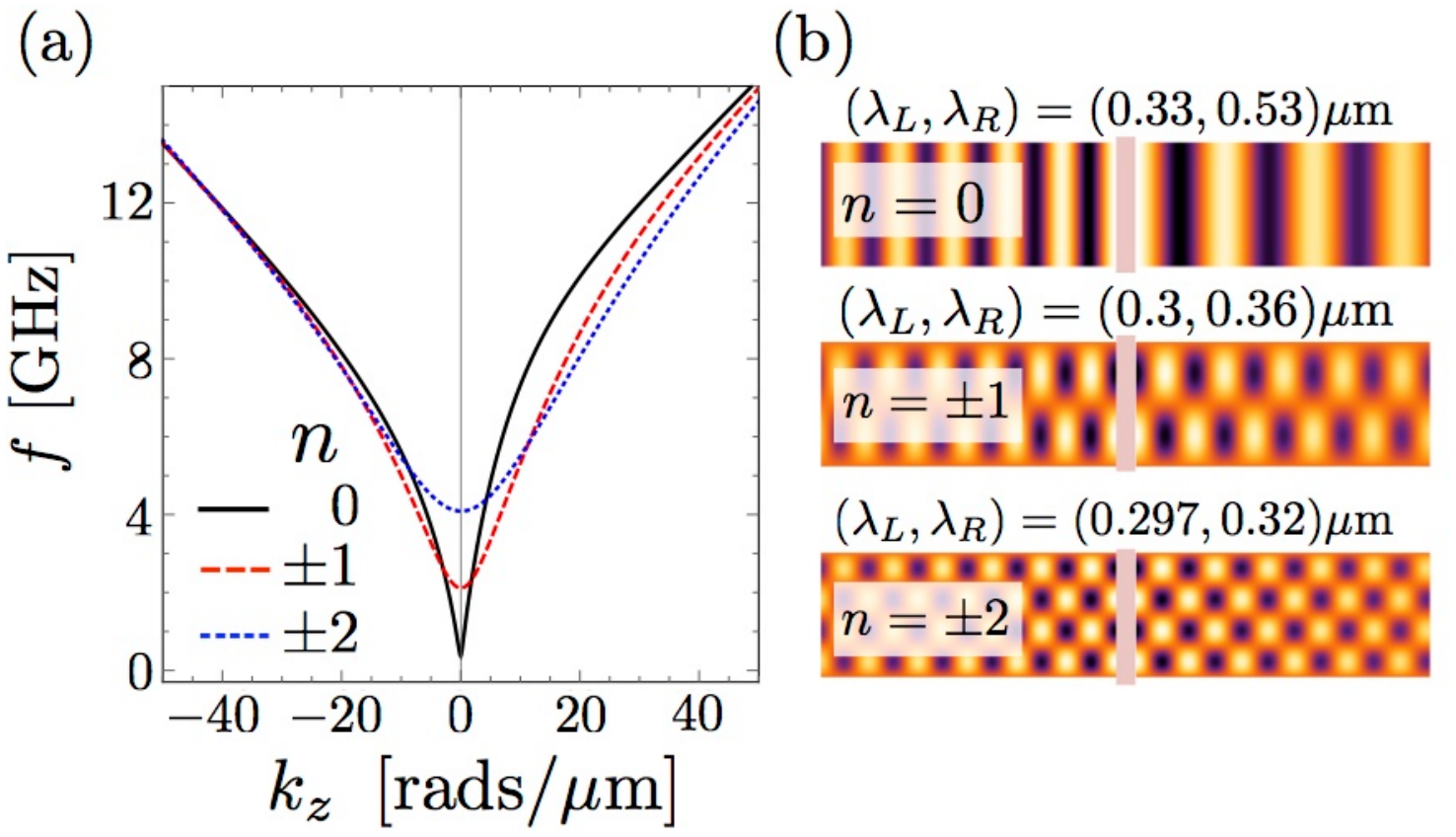}
\end{center}
\caption{(Color online). (a) Eigenfrequency and (b) radial component of SW distribution in an unrolled view of the nanotube at an oscillation frequency of $8$ GHz.  $\lambda_L$ and $\lambda_R$ correspond to the left (L) and right (R) wavelengths of counterpropagating SWs, respectively. An infinitely large permalloy nanotube of outer (inner) radius $R=80$ nm ($r=70$ nm) and under the action of an azimuthal field $\mu_0 H_0=6$ mT is assumed. Curves are presented for the azimuthal wavenumbers $n=0,\pm 1,\pm 2$. }
\label{DispersionRelation}
\end{figure}
%++++++++++++++++++++++++++++++++++++

Figure \ref{DispersionRelation} summarizes our results for SWs with the zeroth-order and the first two higher-order modes in terms of the azimuthal wavenumbers $n=0,\pm 1, \pm 2$. The  dispersion relation $f[n,k_{\text z}]=\omega^n_{k_{\text{z}}}/(2\pi)$ is shown in figure~\ref{DispersionRelation}a). The asymmetry of the dispersion relation is mostly pronounced for the zeroth-order modes, as shown in our previous works~\cite{OtaloraPRL16,OtaloraPRB17}. The mode profiles in an unrolled view of the tube are displayed in figure~ \ref{DispersionRelation}b), where the color code encodes the radial component of the magnetization. These modes excited with an RF field of 8 GHz in the middle of the nanotube (marked with a pink shadowed region) are planar waves with no nodal line for $n=0$, two nodal lines for $n=\pm1$ and four nodal lines for $n=\pm2$. The left and right wavelengths of the counterpropagating waves are denoted as $\lambda_L$ and $\lambda_R$, respectively. 

%+++++++++ Figure3: Simulations +++++++++++
\begin{figure}[th!]
\begin{center}
\includegraphics[width=\linewidth]{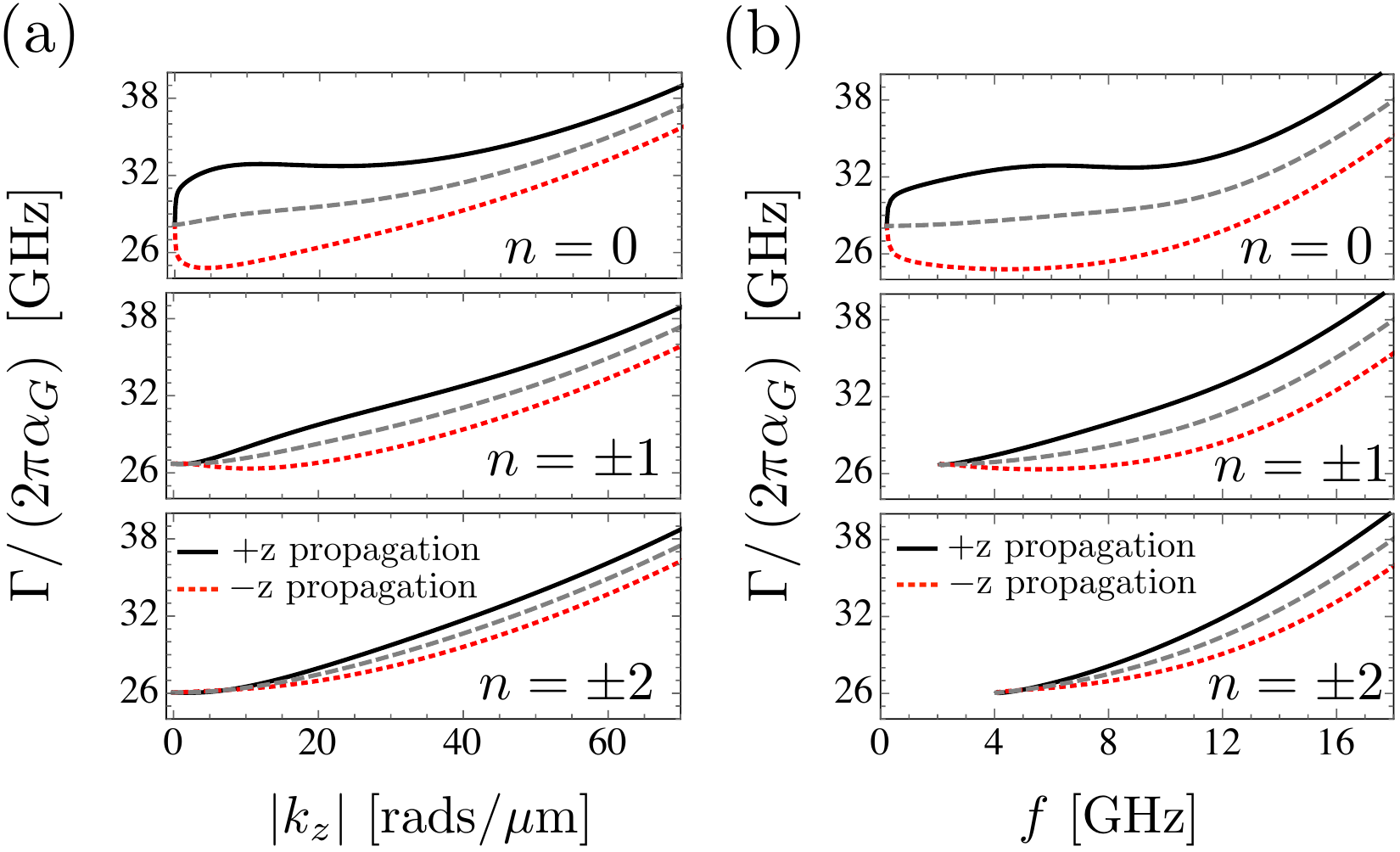}
\end{center}
\caption{(Color online).  Role of curvature in the frequency linewidth of a magnetic nanotube and comparison with the frequency linewidth of a 2D thin film. The frequency linewidth as functions of (a) wavevector  $k_z$ and (b) eigenfrequency  $f$. Curves are presented for the azimuthal wavenumbers $n=0,\pm 1,\pm 2$. The black continuous and red dotted lines correspond to the frequency linewidth as a function of counterpropagating SWs in an infinitely large permalloy nanotube of outer (inner) radius $R=80$ nm ($r=70$ nm) and under the action of an azimuthal field $\mu_0 H_0=6$ mT. The gray dashed lines correspond to the case of an extended 2D thin film with a $10$ nm thickness and with an in-plane magnetization oriented parallel to the magnetic field $\mu_0 H_0=6$ mT. The 2D thin film is assumed to have an in-plane hard axis oriented perpendicularly to $\mu_0 H_0$ and represented by the anisotropic field $\mu_0 H_u=5.9$ mT. The Gilbert damping parameter is denoted by $\alpha_G$. The 2D thin film is supposed to resemble the unrolled view of the nanotubular configuration. The SW wavenumbers $k_z$ and $n$ in the nanotube correspond to the wavevectors  $k_Z=n/\bar \rho$ and $k_X=k_z$ in the thin film, respectively. Therefore, with $\bar\rho=(R+r)/2=75$ nm, the wavevectors $k_Z$ corresponding to $n=0,\pm 1,\pm 2$ are $k_Z= 0,\pm 1/75,\pm 2/75$ nm.
}
\label{Simulations}
\end{figure}
%++++++++++++++++++++++++++++++++++++

The frequency linewidth $\Gamma$ scaled by the Gilbert damping $\alpha_G$ as a function of the SW wavevector $k_z$ and of the eigenfrequency is shown in Figure \ref{Simulations}(a) and (b), respectively. 
The linewidth is asymmetric for the modes with the same wavevector magnitude (see Figure \ref{Simulations}(a)), same eigenfrequency (see Figure \ref{Simulations}(b)) and opposite propagation directions. 
%Since the lifetime is inversely proportional to the linewidth, the counter propagating SWs posses different decay lengths. 
The influence of the nanotube curvature on the frequency linewidth can be better understood by comparing it with the limiting case of a noncurved membrane, which consists of a planar thin film. In this comparison, we consider a planar permalloy thin film with a magnetic configuration similar to that in the unrolled view of the nanotube. Therefore, an extended 2D thin film is assumed with an in-plane applied magnetic field $H_0$ parallel to the homogeneous (in-plane) equilibrium magnetization $\vec \Omega_0$, thickness $d$, and easy-axis anisotropy characterized by the field $H_u$ and  perpendicular to $\vec\Omega_0$. In figure~\ref{Simulations}, we plot with gray dashed lines the frequency linewidth for the planar permalloy thin film, and we plot with black solid and red dotted lines the linewidth for counterpropagating SWs in the NT. These curves are analyzed and discussed in the next section, \textit{Discussions}.
%we plotted in figure~\ref{Simulations} with dashed lines the  frequency dependence of the linewidth for a planar Permalloy thin-film. An extended planar thin-film is assumed with an in-plane applied magnetic field $H_0$ parallel to the homogeneous (in-plane) equilibrium magnetization $\vec \Omega_0$, thickness $d$, and easy-axis anisotropy perpendicular to $\vec\Omega_0$ with the corresponding field $H_u$. 
Note that the range of $\Gamma$ values predicted by our analytical calculations is consistent with the values measured experimentally for permalloy~\cite{TwisselmannJAP03,KalarickalJAP06}. Indeed, assuming typical values for the Gilbert damping parameter $\alpha_G$ between 0.001 and 0.01, one can observe the frequency linewidth ranging between $200$ and $300$ MHz for the given range of eigenfrequencies and wavevectors. 

%+++++++++ Figure4:  +++++++++++
\begin{figure}[th]
\begin{center}
\includegraphics[width=\linewidth]{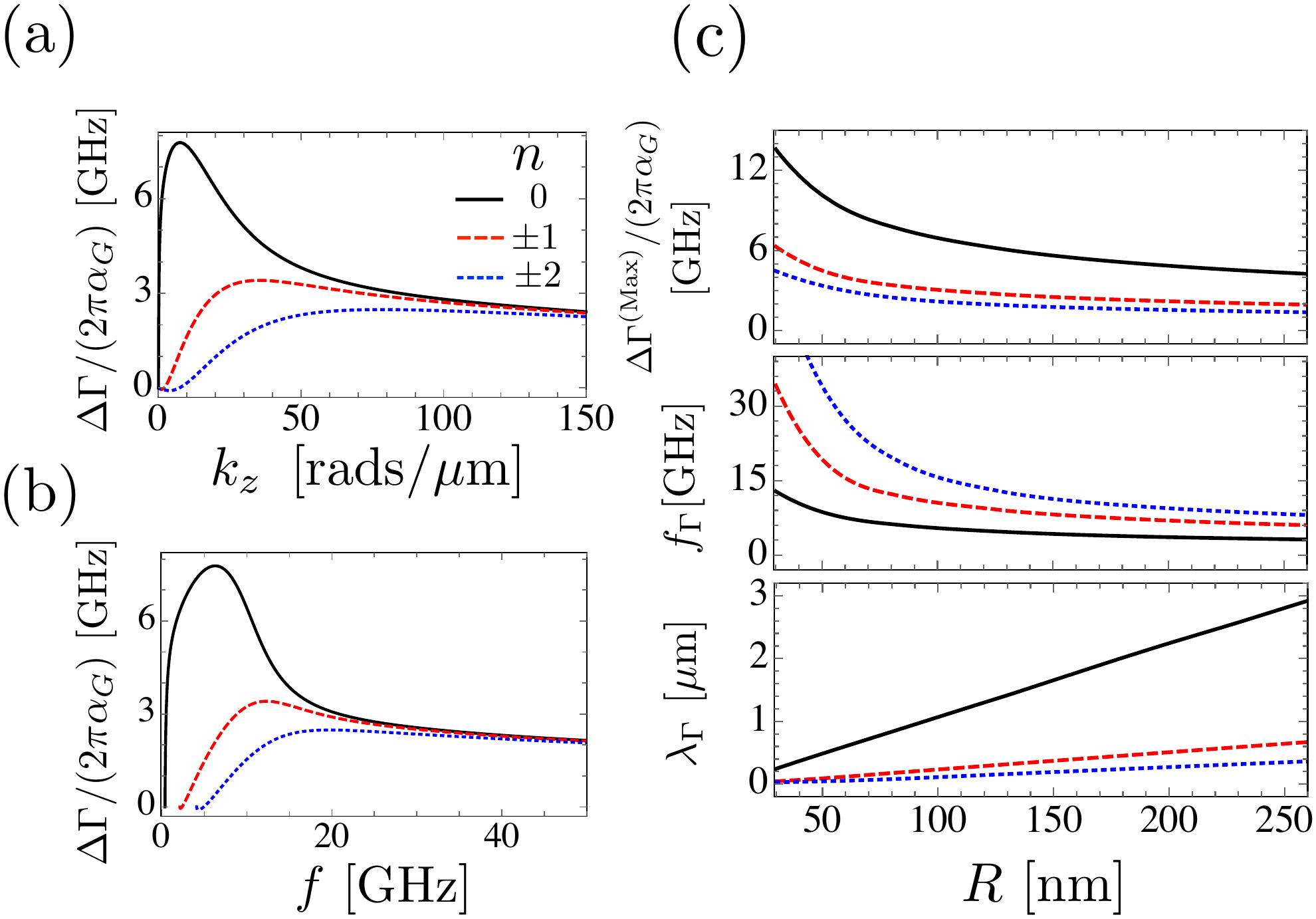}
\end{center}
\caption{(Color online). Frequency linewidth asymmetries (defined by equation~\ref{freqLinewidthAsymmet}) at  the fundamental n = 0 (black lines) and at the first two higher-order azimuthal modes $n = \pm 1$ (red dashed line) and $n=\pm 2$ (blue dotted line). Evolution of the frequency linewidth asymmetry as a function of (a) the wavevector $k_z$ and (b) eigenfrequency of SWs.  (c) Radial dependence of the maximum linewidth asymmetry $\Delta \Gamma^{(\text{Max})}$ (upper curve), of the wavelength $\lambda_{\Gamma}$ (middle curve) and of the eigenfrequency $f_\Gamma$ (bottom figure) of SWs at which the asymmetry reaches its maximum. All cases presented here are calculated for the nanotube thickness $d=10$ nm. An infinitely large nanotube with outer radius $R=80$ nm and an applied circular field of $\mu_0 H_0 = 6$ mT are assumed in (a) and (b). An applied circular field sightly larger than the critical field ($H_0 = 1.01H_c$)\cite{OtaloraJAP15} is assumed for calculating the curves in (c).}
\label{Asymmetries}
\end{figure}
%++++++++++++++++++++++++++++++++++++

The linewidth asymmetry $\Delta \Gamma$, defined as the difference in the linewidth between counterpropagating SWs with the same wavevector magnitude and eigenfrequency, are shown in Figure \ref{Asymmetries} (a) and  (b), respectively. 
As shown, in all modes, $\Delta \Gamma$ shows a maximum defined as $\Delta\Gamma^{(\text{Max})}$, which occurs at a particular frequency ($f=f_\Gamma$) and wavevector $(k_z=k_\Gamma)$.  The evolution of the quantities $\Delta\Gamma^{(\text{Max})}$, $f_\Gamma$ and $\lambda_\Gamma=2\pi/k_\Gamma$  as a function of the NT radius is summarized in Figure \ref{Asymmetries}(c). The linear relation between $\lambda_\Gamma$ and $R$ indicates that the asymmetry is induced by the nanotube mean curvature ($\sim 1/R$). (Note that in the case of the asymmetry in the dispersion relation, a global maximum located at $\lambda_{SW}$ is also found, showing a similar relation with the nanotube mean curvature  ($\sim 1/R$).~\citep{OtaloraPRL16,OtaloraPRB17}) %in a similar fashion than the SWs wave length ($\lambda_{SW}$) and the nanotube radius at which the asymmetry in the dispersion relation is maximum~\citep{OtaloraPRL16,OtaloraPRB17}. This feature can be considered as an evidence of the asymmetry induced by the nanotube mean curvature ($\sim 1/R$).
The changes in the linewidth asymmetries between counterpropagating SWs can be estimated by comparing the values of $\Delta \Gamma^{\text{Max}}$ in figure \ref{Asymmetries}(c) with the values for $\Gamma$ in figure \ref{Simulations}. Indeed, changes of approximately greater than 10\% and up to 20\% for a wide range of nanotube radii between 30 nm and 260 nm are observed, which is comparable with those that can be extracted from experiments on heavy metal/magnetic metal sandwiches wherein linewidth asymmetry results from interfacial DMI.\cite{DiPRL15}.

\begin{figure}[th]
\begin{center}
\includegraphics[width=\linewidth]{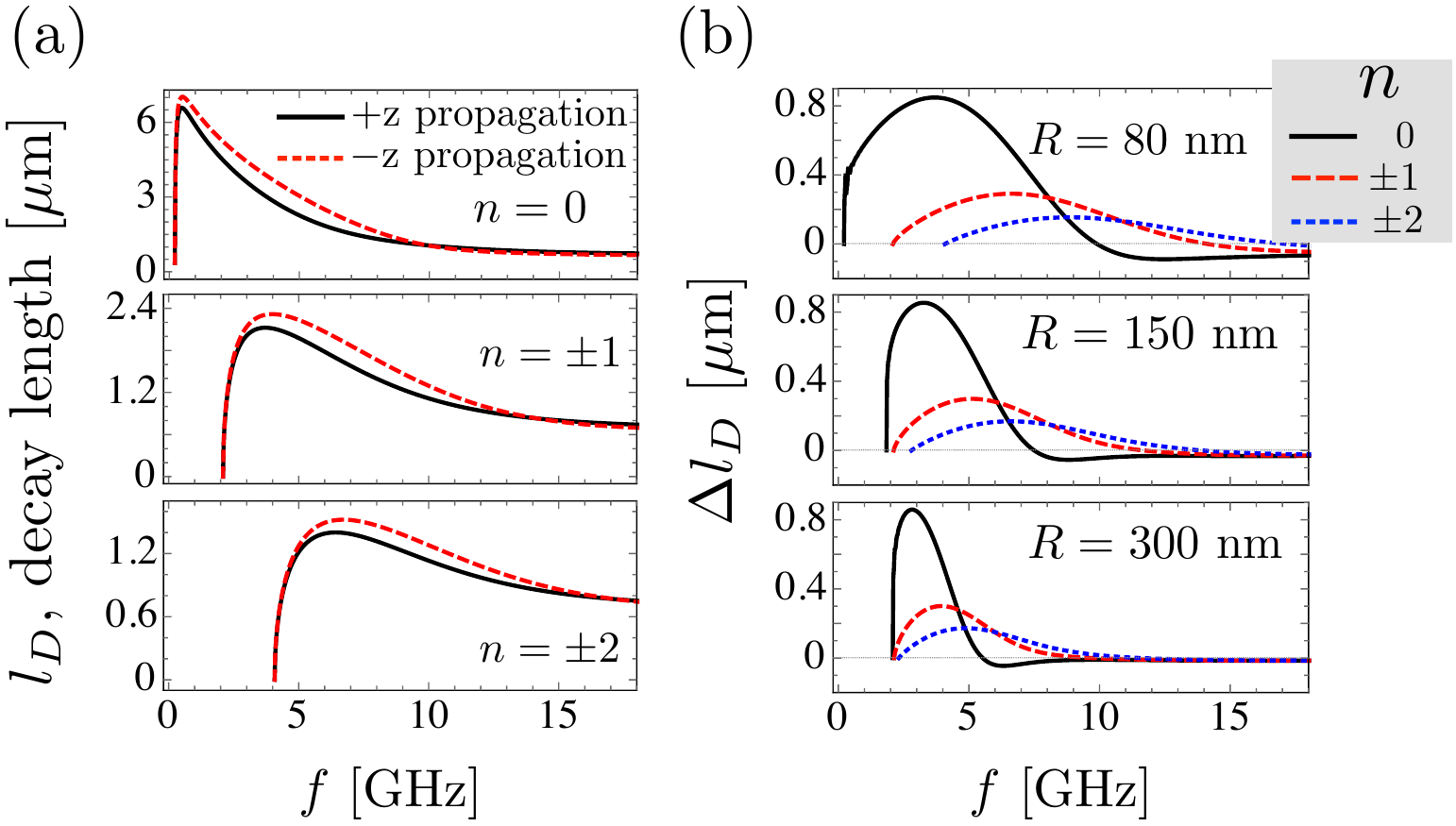}
\end{center}
\caption{(Color online). Decay length of counterpropagating SWs (a) and decay length asymmetries (b) as a function of SW eigenfrequency and for wavenumber $n=0,\pm1,\pm2$. An infinitely large nanotube with outer radius $R=80$ nm, inner radius $r=70$ nm, in a vortex-oriented equilibrium magnetization $\vec M_0=M_s\hat\varphi$ and under the action of an applied circular field $\mu_0 H_0 = 6$ mT is assumed in (a).  The same circular field is assumed for calculating the curves in (b). Note that for all radii in (b), the applied circular field magnitude is larger than the critical field necessary for stabilizing the circular magnetization\cite{OtaloraJAP15}  A Gilbert damping parameter $\alpha_G=0.001$ for the permalloy nanotube was considered.}
\label{DecayLengthAsym}
\end{figure}

The SW decay length, $l_D[n,k_z]\equiv v_g[n,k_z]/\Gamma[n,k_z]$, and the decay length asymmetry $\Delta l_D[n,k_z]=l_D[n,k_z]-l_D[n,-k_z]$ in the nanotubular membrane are shown in figure \ref{DecayLengthAsym}. Here, the group velocity is given by the derivative $v_g[n,k_z]\equiv d\omega_{k_z}^{n}/dk_z$. From figure  \ref{DecayLengthAsym} (a) and (b), we can observe changes ranging between 10\% and  22\% in the decay length in the range of frequencies between 2 GHz and 10 GHz and for three different nanotube radii (these changes are estimated by $\Delta l_D[n,k_z]/l_D[n,k_z]$). Note that the percentage differences are approximately 20\% for $n=0$ and frequencies between 4 GHz and 6 GHz in the case of a nanotube with $R=80$ nm. Moreover,  similar values are also found for the larger nanotube radius $R=300$ nm at the frequency range of 2 - 3 GHz. For larger modes ($n\neq 0$), the percentage difference decreases to approximately 13\% due to the reduced curvature-induced asymmetry of group velocity $v_g$ (the larger is the mode number, the smaller is the difference between the slope of the dispersion relation of counterpropagating SWs) and linewidth $\Gamma$ (reduction of the asymmetries in $\Gamma$ at larger-order mode numbers are explained in the next section). Note that the range of frequencies at which the asymmetries are more evident corresponds to the magneto-chiral dipolar effect induced by the mean nanotube curvature; therefore, it is observed to be more intense around wavevectors $k_z\sim 1/R$.

 %+++++++++ Figure5:  +++++++++++
%++++++++++++++++++++++++++++++++++++

\section{Discussions}
\label{Discussions}

The term responsible for the asymmetry of the frequency linewidth $\Gamma$ (see equation \ref{freqLinewidth}) is $\mathcal{A}_{k_{\text z}}^n$ (originating from the dynamic dipolar field), which also gives rise to the asymmetric dispersion relation shown in ~\textcite{OtaloraPRL16}. This result reveals that nonreciprocities in the frequency linewidth are also induced by the surface curvature. 

The influence of the dynamic dipolar interaction on the linewidth can be better understood if we first discuss the analytical expressions for the thin-film case. Indeed, we present this discussion here for the purpose of completeness and self-containedness of the manuscript. We consider a 2D thin film with the equilibrium magnetic configuration, anisotropies and applied magnetic fields similar to those in the unrolled view of the assumed nanotube configuration. Accordingly, a 2D thin film with in-plane orthonormal axes $\hat X$  and $\hat Z$, out-of-plane axis $\hat Y$, easy anisotropy axis represented by the field $H_u \hat X$, applied magnetic field $H_0 \hat Z$ and equilibrium magnetization $\vec M_0=M_s \hat Z$ will be assumed. This configuration can resemble the unrolled view of the nanotube by mapping the unit vectors $\hat X$, $\hat Y$ and $\hat Z$ to $-\hat z$, $\hat \rho$ and $\hat \varphi$, respectively. As before, $H_u$ is the exchange field arising from the vortex state of the magnetization, and $H_0$ is the applied circular field. The chosen notation for the unit vectors $\hat X$, $\hat Y$ and $\hat Z$ is intentional for a later comparison with the literature. 

According to the literature,\cite{HurbenJAP98,AriasPRB99,LanderosPRB08,GallardoNJP14} the frequency linewidth of an extended thin film $\Gamma_{2D}$ is proportional to the sum of the stiffness fields and can be written as (the film is considered to be free of defects and perfectly smooth, and thus, two-magnon scattering is excluded)

\begin{equation}
\label{PlanarfreqLinewidth}
\Gamma_{2D}[\vec k]=\omega_{M_s} \alpha _G\left(h_X[\vec k]+h_Y[\vec k] \right).
\end{equation}

\noindent Here, $\vec k=k_X \hat X+k_Z\hat Z$ is the in-plane wavevector (out-of-plane wavevector components are not taken into consideration) with $k_X=k \sin\phi_k$ and $k_Z=k\cos\phi_k$ being its components, and  $\phi_{k}$ is the angle between the magnetization and the wavevector.\footnote{Note that the same stiffness fields appear in the dispersion relation ($\omega_{2D}[\vec{k}]=\omega_{M_s}\sqrt{h_X[\vec{k}] h_Y[\vec{k}]}$).} The thin-film thickness (equal to the nanotube thickness) is denoted by $d$, and $h_{X,Y}[\vec k]=H_{X,Y}[\vec k]/M_s$ are the normalized stiffness fields with the following form:

\begin{equation}
\label{StiffnessFields}
\begin{aligned}
H_X[\vec k]&=H_0-H_u+H_{X}^{\text{dip}}[|k|d,\phi_k]+M_sl_{ex}^2k^2,\\
H_Y[\vec k]&=H_0+H_{Y}^{\text{dip}}[|k|d]+M_sl_{ex}^2k^2, \\
\end{aligned}
\end{equation}

\noindent where  $H_{X}^{\text{dip}}[|k|d,\phi_k]$ and $H_{Y}^{\text{dip}}[|k|d]$ are the in-plane and out-of-plane stiffness dipolar fields given as

\begin{equation}
\label{DipoleStiffnessFields}
\begin{aligned}
&H_{X}^{\text{dip}}[|k|d,\phi_k]=M_sF[|k|d]\sin^2[\phi_k],\\
&H_{Y}^{\text{dip}}[|k|d]=M_s(1-F[|k|d]), \\
\end{aligned}
\end{equation}

\noindent with  $F[|k|d]=1-(1-e^{-|k|d})/(|k|d)$. Note that the wavevectors in the 2D thin film $k_X$ and $k_Z$ map respectively to the wavevectors $k_z$ and $\bar k_{\varphi}$ in the nanotube, where $\bar k_{\varphi}=(2\pi/S)\int_r^R k_{\varphi}\rho d\rho=n/\bar \rho$ is the radially averaged azimuthal wavevector $k_{\varphi}=n/\rho$ with $\bar \rho=(R+r)/2$ as the mean nanotube radius. Furthermore, in the nanotube, the angle $\phi_k$ is subtended between the circular equilibrium magnetization $\vec \Omega_0=\hat \varphi$ and the total wavevector $\vec k=k_\varphi\hat\varphi+k_z \hat z$, and it is given by $\tan[\phi_k]=k_z/\bar k_\varphi =\bar \rho k_z/n$. In addition, note that by comparing equation \ref{freqLinewidth} with equation \ref{PlanarfreqLinewidth}, we realize that the quantities $\mathcal{B}_{k_{\text z}}^n$ and $\mathcal{C}_{k_{\text z}}^{n}$ (see equation \ref{StiffnessFieldTube}) play the role of the normalized stiffness fields in the nanotube in an analogous manner to that played by $h_X[\vec k]$ and $h_Y[\vec k]$ in the planar thin film, respectively. The remaining term $\mathcal{A}_{k_{\text z}}^n$ arises from the NT curvature; hence, there is no equivalent term in the planar case unless there is a DMI interaction included, which is not the case in equation \ref{PlanarfreqLinewidth}.

Now, from equations \ref{PlanarfreqLinewidth} and \ref{StiffnessFields}, the frequency linewidth takes the following explicit form:

\begin{equation}
\label{FreqLinewdthkphi}
\begin{aligned}
\Gamma_{2D}[k,\phi_k]=&\alpha_G\frac{\omega_{M_s}}{M_s}\left(2H_0-H_u\right.\\
&\left.+H_{X}^{\text{dip}}[|k|d,\phi_k]+H_{Y}^{\text{dip}}[|k|d]+2M_sl_{ex}^2k^2\right).\\
\end{aligned}
\end{equation}

\noindent This equation is clearly mirror symmetric; by replacing $\phi_k$ with $\phi_k+\pi$, the $\Gamma_{2D}[\vec k]=\Gamma_{2D}[-\vec k]$. The frequency linewidth for the backward volume (BV) ($\Gamma_{2D,BV}$ and $k=k_Z$) and Damon-Eshbach (DE) ($\Gamma_{2D,DE}$ and $k=k_X$) geometries can  be obtained for the angles $\phi_k=0$ and $\phi_k=\pi/2$, respectively. The equations are the following:

\begin{equation}
\label{FreqLinewdthBV}
\begin{aligned}
\Gamma_{2D,BV}=\alpha_G\frac{\omega_{M_s}}{M_s}\big(2H_0-H_u&+H_{Y}^{\text{dip}}[|k_Z|d]\\
&+2M_sl_{ex}^2k_Z^2\big)
\end{aligned}
\end{equation}

\begin{equation}
\label{FreqLinewdthDE}
\begin{aligned}
\Gamma_{2D,DE}&=\alpha_G\frac{\omega_{M_s}}{M_s}\left(2H_0-H_u+M_s+2M_sl_{ex}^2k_X^2\right).\\
\end{aligned}
\end{equation}

Note that the DE and BV geometries in the nanotube can be described from the limiting cases of the intermedium SW configuration (see equation \ref{freqLinewidth}) that is defined by all possible values of $n$ and $k_z$. Analogous with a 2D thin film described by equation \ref{FreqLinewdthkphi}, in the nanotube, it is easily realized that the DE geometry, $\phi_k=\pi/2$, corresponds to $n=0$ and the BV geometry, $\phi_k=0$, to $k_z=0$ and $n\neq0$, or to the limiting case $n\gg \bar\rho k_z$; hence, $k\approx\bar k_\varphi $.

In the following, we would like to focus on the contributions of $H_{X}^{\text{dip}}$ and $H_{Y}^{\text{dip}}$ (see equation \ref{StiffnessFields}) to the frequency linewidth (note that in both fields, the function $F[|k|d]$ introduces the dynamic dipolar field created by the dipolar charges of an SW with wavevector magnitude $k$).  The field $H_{X}^{\text{dip}}$, on the one hand, is nonzero for any angle different than zero $\phi_k\ne0$, and it is a purely dynamic dipolar field that arises from the divergence of the magnetization (volume charges) when an SW is traveling along the $X$ direction. The field $H_{Y}^{\text{dip}}$, on the other hand, is given by the addition of two terms: the static out-of-plane demagnetizing field of an infinite 2D film, $M_s$ (the demagnetizing tensor component $N_Y=1$) and the dynamic field, $-M_s F[|k|d]$, created by the redistribution of surface charges due to the excited SW. 
For $k=0$, the ferromagnetic resonance case (FMR), the $F[|k|d]=0$; therefore, the in-plane component is zero $H_{X}^{\text{dip}}=0$, and the out-of-plane field is equal to the static demagnetizing field, $H_{Y}^{\text{dip}}=M_s$. For nonzero wavevectors, the field $H_{X}^{\text{dip}}$ depends on the angle $\varphi_k$, and the out-of-plane component $H_{Y}^{\text{dip}}$ is reduced by the surface-charge-induced dynamic dipolar field, $-M_sF[|k|d]$. The consequences of $H_{X}^{\text{dip}}$ and $H_{Y}^{\text{dip}}$ (therefore, the role of the dynamic dipolar charges) in the frequency linewidth at the BV, DE and intermedium configurations in a planar thin-film is presented next. 
 
In the BV configuration ($\phi_k = 0$), only the surface dynamic dipolar charges and the out-of-plane field component are nonzero. Consequently, the frequency linewidth will present a dipolar-dominated regime that is introduced by the surface charges via $H_Y^{\text{dip}}$. This results in a decreasing linear dependence on $k_Z$, which is dominant at small wavevector values in $k=k_Z\ll 1/d$ (long-wavelength SWs). However, above a critical wavevector, the exchange field will dominate, and an increasing dependence on $k_Z^2$ is presented. (Note that a similar resemblance in $k_Z$ can be found in the BV dispersion relation).

In the intermedium configuration ($0<\phi_k<\pi/2$), the volume dynamic dipolar field $H_X^{\text{dip}}$ is reduced by a factor of $\sin[\phi_k]^2$; therefore, at a small wavevector magnitude $k\ll 1/d$, the dynamic dipolar contribution from surface charges will dominate on the frequency linewidth (equation \ref{FreqLinewdthkphi}). Therefore, in a similar fashion as in the BV geometry, the frequency linewidth here will present a decreasing linear dependence on $k$, but in a shorter range of wavevector magnitude. This situation can be sightly observed by the gray dashed curve in figure \ref{Simulations}(a) for $n=\pm 1$ at $k_z<3$ rads$/\mu$m (or $k<3$ rads$/\mu$m) and for $n=\pm 2$ at $k_z<5$ rads$/\mu$m (or $k<5$ rads$/\mu$m). 

In the DE configuration ($\phi_k=\pi/2$), the dynamic dipolar fields from both the in-plane and out-of-plane stiffness dipolar fields cancel each other, which is due to the additive contribution of both fields to the linewidth (see equation~\ref{FreqLinewdthDE}). Here, the dynamic surface and volume charges induce in $H_Y^{\text{dip}}$ and $H_X^{\text{dip}}$ the same magnitude (but opposite in sign) dynamic dipolar fields ($M_s F[k_X d]$); hence, the term $M_s F[k_X d]$ in the linewidth cancels out. This means that the effects of volume and surface charges compensate each other, and the dipolar contribution in the frequency linewidth turns into a constant given by the full demagnetizing field ($M_s$). Consequently, in the DE geometry, the wavevector dependence of the linewidth is only driven by the exchange interaction, leading to a $k_X^2$ behavior, as indicated by the gray dashed line in figure~\ref{Simulations}(a) for $n=0$ (or $k_Z=0$ ). 

Note that whether a DE, BV or intermedium configuration,  a linear relation between the frequency linewidth and  SW frequency is generally expected. According to equation \ref{FreqLinewdthkphi}, this can be observed at frequency values where the exchange interaction dominates over the dipolar, shape $H_u$ and Zeeman $H_0$ contributions. Under this condition, the SW frequency and linewidth approximate to $f \approx (\omega_{M_s}/2\pi)l_{ex}^2k^2 $ 
and $\Gamma_{2D}\approx2\alpha_G\omega_{M_s}l_{ex}^2k^2 $, respectively. These two equations lead to $\Gamma_{2D}\approx 4\pi \alpha_G f$, which can be observed by the gray dashed line but at larger frequencies than presented in figure~\ref{Simulations}(b) for $n=0,\pm1,\pm2$.

At this point, we will continue with the analysis of the frequency linewidth equation \ref{freqLinewidth} of the magnetic nanotube. We will utilize the aforementioned SW configurations at the 2D thin film for our next developments. 
%For such a purpose, further similarities between both systems are worth being mentioned. Notice that the quantities $\mathcal{B}_{k_{\text z}}^n$ and $\mathcal{C}_{k_{\text z}}^{n}$ play the role of the normalized stiffness fields in the nanotube in an analogous manner than $h_X[\vec k]$ and $h_Y[\vec k]$ play in the planar thin-film, respectively. The remaining term $\mathcal{A}_{k_{\text z}}^n$ arises from the NT curvature, hence, there is no equivalent term in the planar case unless there is a DMI interaction included, which is not the case in equation \ref{PlanarfreqLinewidth}. 
In the following, the discussion of the frequency linewidth in the tubular membrane will be separated into two subsections: one corresponding to the DE configuration ($n=0$) and the other corresponding to the intermedium configuration ($n\neq0$) that includes the BV geometry at the limit $n\gg \bar\rho k_z$.

\subsection{Linewidth in DE configuration ($n=0$)}

As  previously mentioned, the asymmetries in the frequency linewidth are given by $\mathcal{A}_{k_{\text z}}^n$. This term is linear in $k_z$ with a positive slope for wavevector values up to $k_z\sim1/R$. The term $\mathcal{A}_{k_{\text z}}^n$ is responsible for the differences in the linewidth between the nanotube and the 2D thin film at the DE configuration: in the nanotube for SWs traveling with $k_z >0$  ($+$ z propagation), the linewidth increases linearly, and for the oppositely propagating SWs $k_z <0$  ($-$ z propagation), the linewidth decreases linearly. These features are observed in figure \ref{Simulations}(a) for $n=0$ and $|k_z|<4$ rads$/\mu$m$^{-1}$. A similar linear behavior between the frequency linewidth and eigenfrequency of counterpropagating SWs at $f< 1$ GHz can also be observed in figure \ref{Simulations}(b). This result can be understood from the first-order expansion of the eigenfrequency in terms of the wavevector, $f=f_0+\nu_0 k_z$, where $\nu_0$ is a constant slope that depends on the nanotube radius and $f_0$ is the nanotube ferromagnetic resonance - FMR.\cite{OtaloraPRB17} Since $\Gamma[n=0,k_z]$ and $f$  are linear in $k_z$ at first-order expansion,  $\Gamma$ is also linear in $f$ at the same expansion order. From this linear relation,  the not mirror symmetry of the frequency linewidth $\Gamma[n=0,k_z]\neq\Gamma[n=0,-k_z]$ is  clearly realized. The linewidth asymmetry with a decreasing tendency can also be obtained at the exchange-dominated regime: in the range $k_z\gg 1/R$, the asymmetry will eventually be neglected for sufficiently large $k_z$. Moreover, as in the 2D thin film, in the nanotube, the exchange-dominated frequency linewidth is also linear in the eigenfrequency $f$, a situation that can be obtained at larger frequencies than  presented in figure \ref{Simulations}(b)

The frequency linewidth asymmetry mentioned above can be understood in terms of the dynamic dipolar field created by the dynamic dipolar charges. In contrast to the preserved mirror symmetry of dipolar charges in 2D thin films, in the nanotube, the symmetry is disturbed by the tubular mean curvature. In the nanotube, the surface charges conserve the mirror symmetry, and the magnitude of the volume charges violate it.~\cite{OtaloraPRL16} This breaks the mirror symmetry of the dynamic dipolar fields and consequently violates the left-right chiral symmetry of the frequency linewidth of counterpropagating SWs. On the one hand, for $k_z<0$, the divergence of the dynamic magnetization and the strength of the volume dynamic dipolar field are reduced. Hence, the field created by surface charges dominates. Understanding that the surface charges and its dynamic dipolar field strength decrease with increasing the wavevector magnitude $k_z$, it is found that in a first-order expansion in $k_z$, the frequency linewidth shows a non-negligible decreasing tendency. Note that this situation resembles a behavior similar to that in the BV geometry (or in the intermedium geometry) in 2D thin films, wherein the effects of the volume dynamic charges are absent (or reduced).  On the other hand, for SWs with $k_z>0$, the divergence of the dynamic magnetization and the volume dynamic charges are enhanced. Thus, the dynamic dipolar fields arising from the volume charges dominate over the dynamic fields from the surface charges. Realizing that the divergence of the SW distribution and the volume dynamic dipolar field increase with increasing  SW wavevector, it is found that the frequency linewidth has a non-negligible linear increasing tendency with $k_z$. The aforementioned linear trend between $\Gamma$ and $k_z$ is reflected in figure~\ref{Simulations}(a) at small values of $k_z$, as previously mentioned (for comparison, note that in the 2D thin film, both dynamic fields balance each other, leading to an absent linear $k_z$ contribution in the frequency linewidth). Finally, for a larger wavevector magnitude, the exchange interaction will eventually dominate over the dipolar interaction, and the frequency linewidth will take a $k_z^2$ dependence, as discussed previously. (Recall that in a 2D thin film in a DE geometry, the total dynamic dipolar field contribution to the linewidth is constant and equal to the magnetostatic demagnetizing field, leading to a $k_z^2$ quadratic dependence in the frequency linewidth that originates from the exchange energy).

\subsection{Linewidth in intermedium configuration ($n\neq0$)}
For higher-order modes ($n\neq 0$), the geometry of SWs is defined by the angle $\phi_k$ subtended between the wavevector $\vec k=k_z\hat z+\bar{k}_{\varphi}\hat \varphi$ (where $\bar{k}_{\varphi}=n/\bar \rho$) and the equilibrium magnetization $\vec \Omega_0$. The BV geometry, $\phi_k=\pi/2$, is characterized by the lack of asymmetries in the frequency linewidth and is given by whether doing $k_z=0$ or in the limit of $|n|\gg \bar \rho |k_z|\sim R |k_z|$. As observed in figure \ref{Simulations}(a) for $n=\pm 1,\pm2$, the absence of asymmetries in the case of $k_z=0$ can be understood if we take into account that the dynamic volume charges (obtained from the divergence of the dynamic excitation - the SW) are free of curvature-induced asymmetries. Similarly, no asymmetries must also be observed in terms of the eigenfrequency $f$, as is indeed shown in figure \ref{Simulations}(b) for $n=\pm 1,\pm2$ at the eigenfrequency $f[k_z=0]$ where all curves converge. More details about $k_z=0$ will be presented in subsection \ref{cerokz}. In the other BV geometry case, at the limit $|n|\gg \bar \rho |k_z|$, asymmetries in the frequency linewidth are not presented. For any value of $k_z$ (large $|k_z|\gg1/R$, small $|k_z|\ll 1/R$ or located at the optimal asymmetry $|k_z|\sim 1/R$), the integer nature of $n$ leads to the limit $|n|\gg 0$ as the sufficient condition to define the BV geometry. For illustration, we consider the range $|k_z|\sim 1/R$. Accordingly, we obtain $\bar \rho |k_z|\sim 1$; therefore, $|n|\gg1$. In terms of the azimuthal wavevector, the condition $|n|\gg1$ means that $|k_\varphi|\gg 1/R$. Consequently, the total wavevector magnitude is $k\approx |k_\varphi|\gg 1/R$, which is located at the exchange-dominated regime that is characterized for no curvature-induced asymmetries (via dipolar interaction) in the frequency linewidth. 

For finite values of $n$, we obtain the intermedium configuration defined by $0<\phi_k<\pi/2$. In this case, less intense asymmetries in the frequency linewidth than with $n=0$ appear, which is a consequence of the fact that the larger is the value of $n$, the more important is the exchange interaction. Figure \ref{Simulations} shows reduced asymmetric frequency linewidth of counterpropagating SWs for $n=\pm1,\pm 2$ in a wide range of wavevectors $k_z$ and eigenfrequencies $f$, and the asymmetries are quantified in figure \ref{Asymmetries}. Since this nonreciprocal effect is curvature induced, the asymmetry $\Delta \Gamma$ must decrease (i) by increasing the wavevector $k_z$ at $k_z>1/R\approx 12.5$ rads$/\mu$m (see figure \ref{Asymmetries}(a)) and consequently (ii) by increasing the SW frequencies for over $f[1/R]\approx 5.6$ GHz (see figure \ref{Asymmetries}(b)). These values correspond to a nanotube with $R=80$ nm. An analytical expression of the frequency linewidth for finite $n$ and small wavevector range $k_z\ll 1/R$ can be obtained, showing a direct resemblance with 2D thin films with DMI; nevertheless, this case is presented in subsection \ref{smallkz}.

\section{LIMITING CASES}
\label{LimitCases}

For completeness of our comparative analysis between the 2D thin film and nanotubular membrane, a few limiting cases of the nanotube frequency linewidth are presented. In the following, we will address three cases consisting of A. $k_z =0$, B. $k_z \gg 1/R$, and C. $kz\neq 0$ ($k_{\text z}\ll 1/R$), in which the nanotubular curvature can be neglected; therefore, the frequency linewidth is reduced to well-known expressions for 2D films. 

\subsection{\textbf{For $k_z =0$}}
\label{cerokz}
In this case, SWs with a homogeneous profile along the nanotube long axis propagate only in the azimuthal direction with wavenumber $n$. Since the propagation direction and the magnetization are parallel, this case resembles the BV geometry of the SW propagation in planar thin films. The frequency linewidth takes the form

\begin{equation}
\label{freqLinekz0}
\Gamma(n,0)= \alpha _G\omega_{M_s}\left((2n^2-1)h_u+2h_0+\mathcal{J}_0^n\right),
\end{equation}

\noindent which for the exchange-dominated regime, i.e., $n\gg 1$, is written as

\begin{equation}
\label{freqLinekz0ngg1-0}
\Gamma(n\gg 1,0)=\omega_{M_s} \alpha _G\left(2 n^2 h_u+2h_0+\frac{1}{n}\frac{R^2+r^2}{R^2-r^2}\right).
\end{equation}

\noindent For nanotubes with a radius much larger than the thickness $R\gg d$, the previous equation can be written as

\begin{equation}
\label{freqLinekz0ngg1}
\Gamma(n\gg 1,0)=\omega_{M_s} \alpha _G\left(2 l_{ex}^2 \bar k_{\varphi}^2+2h_0+\frac{1}{\bar k_{\varphi} d}\right),
\end{equation}

\noindent where $d$ is the nanotube thickness. Equation \ref{freqLinekz0ngg1} is similar to the expression of the linewidth that one can obtain for 2D thin films for the BV modes. %(use planar thin-film frequency linewidth equation $\left(\frac{\Lambda(\vec k)}{\Omega-\Omega(\vec k)}\right)_{\Omega=\Omega(\vec k)}$ with $\Lambda(\vec k)$ and $\Omega(\vec k)$ defined in page 8, first paragraph and assuming BV-modes - i.e.  $\phi_k=\pi/2$ - in conjunction with equations (18) and (19) in Ref. \citep{GallardoNJP14}). 
In the case of zero azimuthal wavenumber $n=0$ (typically called the ferromagnetic resonance (FMR) mode), the frequency linewidth is 

\begin{equation}
\label{freqLinekz0n0}
\Gamma(0,0)=\alpha _G\omega_{M_s}\left(2h_0-h_u+1\right),
\end{equation}

\noindent which resembles the FMR frequency linewidth known from the Kittel formula\citep{Kittel:ISSP} for a thin film with a homogeneous in-plane magnetization parallel to the applied magnetic field $H_0$.

\subsection{\textbf{For $k_z \gg 1/ R$}}
\label{largekz}
In this case, the SW wavelength is considerably smaller than the nanotube radius. Therefore, the SWs are exchange dominated, and the dispersion relation is symmetric. 
The frequency linewidth can be written as

\begin{equation}
\begin{aligned}
\label{freqLinekzgg1n}
\Gamma(n,k_z\gg 1/R)= \alpha _G\omega_{M_s}&\left(2 l_{ex}^2 k_{z}^2\right.\\
 &\left. +(2n^2-1)h_u+2h_0+1\right). 
\end{aligned}
\end{equation}

For $n=0$ or SWs with no nodal lines, equation~\ref{freqLinekzgg1n} resembles the linewidth of Damon-Eshbach (DE) modes in 2D thin films with in-plane magnetization (see equation \ref{FreqLinewdthDE}).  

\subsection{\textbf{For $k_z \neq0$} ($k_z~\ll~1/R$)}
\label{smallkz}
As the last case, the linewidth of SWs with small wavevectors is discussed. At this limit, the hyperfunctions shown in equation \ref{HyperfuncDef} can be approximated as shown in ~\textcite{OtaloraPRB17}, resulting in $\mathcal{A}_{k_z}^0\approx \nu_n k_{z}$, $\mathcal{J}_{k_z}^0\approx (1-u_n) k_{z}^2$ and $\mathcal{L}_{k_z}^0\approx (1-\textsl{w}_n) k_{z}^2$. Therefore, the frequency linewidth will have the following complex expression:

\begin{widetext}
\begin{equation}
\label{freqLinekzll1n}
\begin{aligned}
\Gamma(n,k_z\ll 1/R)\approx\alpha _G \omega_{M_s} &\left((2l_{ex}^2+\textsl{w}_n-u_n)k_z^2+(2n^2-1)h_u+2h_0+1\right)\times\\
 &\left(1+\frac{\nu_n k_z}{\sqrt{((l_{ex}^2+\textsl{w}_n)k_z^2+(n^2-1)h_u+h_0)((l_{ex}^2-u_n)k_z^2+n^2 h_u+h_0+1)}}\right)
\end{aligned}
\end{equation}
\end{widetext}

\noindent where $\nu_n$, $u_n$ and $\textsl{w}_n$ can only be calculated numerically (for details, see Ref.~\cite{OtaloraPRB17}). However, these coefficients were already calculated for $n=0$ (see figure 7 in Ref. \citep{OtaloraPRB17}), and for this particular case, the frequency linewidth takes the form

\begin{widetext}
\begin{equation}
\label{freqLinekzll1n0}
\begin{aligned}
\Gamma(0,k_z\ll 1/R)\approx\alpha _G&\omega_{M_s} \left((2l_{ex}^2+\textsl{w}-u)k_z^2-h_u+2h_0+1\right)\times\\
 &\left(1+\frac{\nu_0 k_z}{\sqrt{((l_{ex}^2+\textsl{w}_0)k_z^2-h_u+h_0)((l_{ex}^2-u_0)k_z^2+h_0+1)}}\right).
\end{aligned}
\end{equation}
\end{widetext}

Equation~\ref{freqLinekzll1n0} shows that the asymmetry is linear in $k_z$ for counterpropagating SWs.%Editor: Please ensure that the intended meaning has been maintained in the above edit.
 It can be observed by taking into account the change in the sign of $k_z$ for opposite propagations. Note that the curvature-induced asymmetries in the frequency linewidth are similar to those obtained for heavy metal/magnetic metal 2D sandwiches with interfacial DMI for SWs in the DE configuration. In Ref.~\cite{DiPRL15}, a similar linear term, originating from the interfacial DMI, appears in the frequency linewidth expression.

\section{Conclusions} 
\label{Conclusions}
We have elaborated an analytical model of the frequency linewidth for curved magnetic membranes with a tubular geometry. It is found that the linewidth is asymmetric regarding the eigenfrequency of the SWs and the sign of the propagation vector. The asymmetry originates from the dynamic dipolar fields and is directly related to the mean curvature of the magnetic nanotube. In contrast to the 2D thin film, for SWs in the DE geometry, the dynamic dipolar fields produce an extra contribution that is linear in $k$ in the linewidth. Consequently, SWs propagating to opposite directions but with the same frequency have different lifetimes, group velocities and thus different decay lengths. The predicted asymmetry of the linewidth shows a similar tendency as that reported for the dispersion relation of the nanotubes in our previous works~\citep{OtaloraPRL16,OtaloraPRB17}. We also show that the asymmetries presented in the group velocity and frequency linewidth lead to a decay length with evident asymmetries, which is also calculated. Indeed, changes between 10\% and 20\% in frequency linewidth and between 10\% and 22\% in the decay length for counterpropagating SWs are predicted. Furthermore, it is shown that for modes with azimuthal wavevector only ($k_z=0$) and large radius, the expression of the linewidth reduced to a formula similar to that of the BV modes in planar thin films. In the limiting case of SWs with a wavelength that is considerably smaller than the nanotube radius ($k_z \gg 1/ R$) and no nodal lines (higher-order modes excluded: $n=0$), the linewidth resembles the linewidth that one obtains for the DE modes in planar thin films. Finally, in the last case for intermediate SW wavelengths $k_z \neq0$ ($k_z~\ll~1/R$), the obtained expression for the linewidth is similar to that obtained for heavy metal/magnetic metal 2D sandwiches with interfacial DMI for SWs in the DE configuration. However, the linear term responsible for the asymmetry originates from the dynamic volume changes. We believe that our findings represent a step forward toward the realization of 3D curvilinear magnonic devices.

\section*{Acknowledgments}
Fruitful discussions with Kai Wagner are acknowledged. 

\section*{Author contributions}

J.A.O. developed the analytical model and wrote the manuscript. All authors interpreted and discussed the results and co-wrote the manuscript.

\section*{Additional information}
The authors declare no competing financial interests. Reprints and permission information is available online at http://XXXXXX. Correspondence and requests for materials should be addressed to J.A.O.

%%%%%%%%%%%%%%%%%%%%%%%%%%%%%%%%%%%%%%%%%%%%%%%%%%%%%%%%%%%%%%%%%%%%%
%% The appropriate \bibliographystyle and \bibliography commands
%% should be placed here.
%%%%%%%%%%%%%%%%%%%%%%%%%%%%%%%%%%%%%%%%%%%%%%%%%%%%%%%%%%%%%%%%%%%%%
\section*{References}
%\bibliography{TailoringRef}

\begin{thebibliography}{32}%
\makeatletter
\providecommand \@ifxundefined [1]{%
 \@ifx{#1\undefined}
}%
\providecommand \@ifnum [1]{%
 \ifnum #1\expandafter \@firstoftwo
 \else \expandafter \@secondoftwo
 \fi
}%
\providecommand \@ifx [1]{%
 \ifx #1\expandafter \@firstoftwo
 \else \expandafter \@secondoftwo
 \fi
}%
\providecommand \natexlab [1]{#1}%
\providecommand \enquote  [1]{``#1''}%
\providecommand \bibnamefont  [1]{#1}%
\providecommand \bibfnamefont [1]{#1}%
\providecommand \citenamefont [1]{#1}%
\providecommand \href@noop [0]{\@secondoftwo}%
\providecommand \href [0]{\begingroup \@sanitize@url \@href}%
\providecommand \@href[1]{\@@startlink{#1}\@@href}%
\providecommand \@@href[1]{\endgroup#1\@@endlink}%
\providecommand \@sanitize@url [0]{\catcode `\\12\catcode `\$12\catcode
  `\&12\catcode `\#12\catcode `\^12\catcode `\_12\catcode `\%12\relax}%
\providecommand \@@startlink[1]{}%
\providecommand \@@endlink[0]{}%
\providecommand \url  [0]{\begingroup\@sanitize@url \@url }%
\providecommand \@url [1]{\endgroup\@href {#1}{\urlprefix }}%
\providecommand \urlprefix  [0]{URL }%
\providecommand \Eprint [0]{\href }%
\providecommand \doibase [0]{http://dx.doi.org/}%
\providecommand \selectlanguage [0]{\@gobble}%
\providecommand \bibinfo  [0]{\@secondoftwo}%
\providecommand \bibfield  [0]{\@secondoftwo}%
\providecommand \translation [1]{[#1]}%
\providecommand \BibitemOpen [0]{}%
\providecommand \bibitemStop [0]{}%
\providecommand \bibitemNoStop [0]{.\EOS\space}%
\providecommand \EOS [0]{\spacefactor3000\relax}%
\providecommand \BibitemShut  [1]{\csname bibitem#1\endcsname}%
\let\auto@bib@innerbib\@empty
%</preamble>
\bibitem [{\citenamefont {Bloch}(1930)}]{BlochZP30}%
  \BibitemOpen
  \bibfield  {author} {\bibinfo {author} {\bibfnamefont {F.}~\bibnamefont
  {Bloch}},\ }\href@noop {} {\bibfield  {journal} {\bibinfo  {journal}
  {Zeitschrift f{\"u}r Physik}\ }\textbf {\bibinfo {volume} {61}},\ \bibinfo
  {pages} {206} (\bibinfo {year} {1930})}\BibitemShut {NoStop}%
\bibitem [{\citenamefont {Grundler}(2015)}]{GRUNDNPHYS15}%
  \BibitemOpen
  \bibfield  {author} {\bibinfo {author} {\bibfnamefont {D.}~\bibnamefont
  {Grundler}},\ }\href {http://dx.doi.org/10.1038/nphys3349} {\bibfield
  {journal} {\bibinfo  {journal} {Nat Phys}\ }\textbf {\bibinfo {volume}
  {11}},\ \bibinfo {pages} {438} (\bibinfo {year} {2015})}\BibitemShut
  {NoStop}%
\bibitem [{\citenamefont {Grundler}(2016)}]{GrundlerNatNan16}%
  \BibitemOpen
  \bibfield  {author} {\bibinfo {author} {\bibfnamefont {D.}~\bibnamefont
  {Grundler}},\ }\href {http://dx.doi.org/10.1038/nnano.2016.16} {\bibfield
  {journal} {\bibinfo  {journal} {Nature nanotechnology}\ } (\bibinfo {year}
  {2016})}\BibitemShut {NoStop}%
\bibitem [{\citenamefont {Chumak}\ \emph
  {et~al.}(2015{\natexlab{a}})\citenamefont {Chumak}, \citenamefont
  {Vasyuchka}, \citenamefont {Serga},\ and\ \citenamefont
  {Hillebrands}}]{Chumak14}%
  \BibitemOpen
  \bibfield  {author} {\bibinfo {author} {\bibfnamefont {A.~V.}\ \bibnamefont
  {Chumak}}, \bibinfo {author} {\bibfnamefont {V.~I.}\ \bibnamefont
  {Vasyuchka}}, \bibinfo {author} {\bibfnamefont {A.~A.}\ \bibnamefont
  {Serga}}, \ and\ \bibinfo {author} {\bibfnamefont {B.}~\bibnamefont
  {Hillebrands}},\ }\href {http://dx.doi.org/10.1038/nphys3347} {\bibfield
  {journal} {\bibinfo  {journal} {Nat Phys}\ }\textbf {\bibinfo {volume}
  {11}},\ \bibinfo {pages} {453} (\bibinfo {year}
  {2015}{\natexlab{a}})}\BibitemShut {NoStop}%
\bibitem [{\citenamefont {Vogt}\ \emph {et~al.}(2012)\citenamefont {Vogt},
  \citenamefont {Schultheiss}, \citenamefont {Jain}, \citenamefont {Pearson},
  \citenamefont {Hoffmann}, \citenamefont {Bader},\ and\ \citenamefont
  {Hillebrands}}]{Vogt_APL12}%
  \BibitemOpen
  \bibfield  {author} {\bibinfo {author} {\bibfnamefont {K.}~\bibnamefont
  {Vogt}}, \bibinfo {author} {\bibfnamefont {H.}~\bibnamefont {Schultheiss}},
  \bibinfo {author} {\bibfnamefont {S.}~\bibnamefont {Jain}}, \bibinfo {author}
  {\bibfnamefont {J.~E.}\ \bibnamefont {Pearson}}, \bibinfo {author}
  {\bibfnamefont {A.}~\bibnamefont {Hoffmann}}, \bibinfo {author}
  {\bibfnamefont {S.~D.}\ \bibnamefont {Bader}}, \ and\ \bibinfo {author}
  {\bibfnamefont {B.}~\bibnamefont {Hillebrands}},\ }\href {\doibase
  http://dx.doi.org/10.1063/1.4738887} {\bibfield  {journal} {\bibinfo
  {journal} {Applied Physics Letters}\ }\textbf {\bibinfo {volume} {101}},\
  \bibinfo {eid} {042410} (\bibinfo {year} {2012}),\
  http://dx.doi.org/10.1063/1.4738887}\BibitemShut {NoStop}%
\bibitem [{\citenamefont {Chumak}\ \emph
  {et~al.}(2014{\natexlab{a}})\citenamefont {Chumak}, \citenamefont {Serga},\
  and\ \citenamefont {Hillebrands}}]{Chumak14MT}%
  \BibitemOpen
  \bibfield  {author} {\bibinfo {author} {\bibfnamefont {A.~V.}\ \bibnamefont
  {Chumak}}, \bibinfo {author} {\bibfnamefont {A.~A.}\ \bibnamefont {Serga}}, \
  and\ \bibinfo {author} {\bibfnamefont {B.}~\bibnamefont {Hillebrands}},\
  }\href {http://dx.doi.org/10.1038/ncomms5700} {\bibfield  {journal} {\bibinfo
   {journal} {Nat Commun}\ }\textbf {\bibinfo {volume} {5}} (\bibinfo {year}
  {2014}{\natexlab{a}})}\BibitemShut {NoStop}%
\bibitem [{\citenamefont {Vogt}\ \emph {et~al.}(2014)\citenamefont {Vogt},
  \citenamefont {Fradin}, \citenamefont {Pearson}, \citenamefont {Sebastian},
  \citenamefont {Bader}, \citenamefont {Hillebrands}, \citenamefont
  {Hoffmann},\ and\ \citenamefont {Schultheiss}}]{Vogt14}%
  \BibitemOpen
  \bibfield  {author} {\bibinfo {author} {\bibfnamefont {K.}~\bibnamefont
  {Vogt}}, \bibinfo {author} {\bibfnamefont {F.~Y.}\ \bibnamefont {Fradin}},
  \bibinfo {author} {\bibfnamefont {J.~E.}\ \bibnamefont {Pearson}}, \bibinfo
  {author} {\bibfnamefont {T.}~\bibnamefont {Sebastian}}, \bibinfo {author}
  {\bibfnamefont {S.~D.}\ \bibnamefont {Bader}}, \bibinfo {author}
  {\bibfnamefont {B.}~\bibnamefont {Hillebrands}}, \bibinfo {author}
  {\bibfnamefont {A.}~\bibnamefont {Hoffmann}}, \ and\ \bibinfo {author}
  {\bibfnamefont {H.}~\bibnamefont {Schultheiss}},\ }\href
  {http://dx.doi.org/10.1038/ncomms4727} {\bibfield  {journal} {\bibinfo
  {journal} {Nat Commun}\ }\textbf {\bibinfo {volume} {5}} (\bibinfo {year}
  {2014})}\BibitemShut {NoStop}%
\bibitem [{\citenamefont {Lenk}\ \emph {et~al.}(2011)\citenamefont {Lenk},
  \citenamefont {Ulrichs}, \citenamefont {Garbs},\ and\ \citenamefont
  {M{\"u}nzenberg}}]{Lenk11}%
  \BibitemOpen
  \bibfield  {author} {\bibinfo {author} {\bibfnamefont {B.}~\bibnamefont
  {Lenk}}, \bibinfo {author} {\bibfnamefont {H.}~\bibnamefont {Ulrichs}},
  \bibinfo {author} {\bibfnamefont {F.}~\bibnamefont {Garbs}}, \ and\ \bibinfo
  {author} {\bibfnamefont {M.}~\bibnamefont {M{\"u}nzenberg}},\ }\href@noop {}
  {\bibfield  {journal} {\bibinfo  {journal} {Phys. Rep.}\ }\textbf {\bibinfo
  {volume} {507}},\ \bibinfo {pages} {107} (\bibinfo {year}
  {2011})}\BibitemShut {NoStop}%
\bibitem [{\citenamefont {Ot{\'a}lora}\ \emph {et~al.}(2015)\citenamefont
  {Ot{\'a}lora}, \citenamefont {Cort{\'e}s-Ortu{\~n}o}, \citenamefont
  {G{\"o}rlitz}, \citenamefont {Nielsch},\ and\ \citenamefont
  {Landeros}}]{OtaloraJAP15}%
  \BibitemOpen
  \bibfield  {author} {\bibinfo {author} {\bibfnamefont {J.}~\bibnamefont
  {Ot{\'a}lora}}, \bibinfo {author} {\bibfnamefont {D.}~\bibnamefont
  {Cort{\'e}s-Ortu{\~n}o}}, \bibinfo {author} {\bibfnamefont {D.}~\bibnamefont
  {G{\"o}rlitz}}, \bibinfo {author} {\bibfnamefont {K.}~\bibnamefont
  {Nielsch}}, \ and\ \bibinfo {author} {\bibfnamefont {P.}~\bibnamefont
  {Landeros}},\ }\href@noop {} {\bibfield  {journal} {\bibinfo  {journal}
  {Journal of Applied Physics}\ }\textbf {\bibinfo {volume} {117}},\ \bibinfo
  {pages} {173914} (\bibinfo {year} {2015})}\BibitemShut {NoStop}%
\bibitem [{\citenamefont {Chumak}\ \emph
  {et~al.}(2014{\natexlab{b}})\citenamefont {Chumak}, \citenamefont {Serga},\
  and\ \citenamefont {Hillebrands}}]{ChumakNCOM14}%
  \BibitemOpen
  \bibfield  {author} {\bibinfo {author} {\bibfnamefont {A.~V.}\ \bibnamefont
  {Chumak}}, \bibinfo {author} {\bibfnamefont {A.~A.}\ \bibnamefont {Serga}}, \
  and\ \bibinfo {author} {\bibfnamefont {B.}~\bibnamefont {Hillebrands}},\
  }\href@noop {} {\bibfield  {journal} {\bibinfo  {journal} {Nature
  communications}\ }\textbf {\bibinfo {volume} {5}} (\bibinfo {year}
  {2014}{\natexlab{b}})}\BibitemShut {NoStop}%
\bibitem [{\citenamefont {Chumak}\ \emph
  {et~al.}(2015{\natexlab{b}})\citenamefont {Chumak}, \citenamefont
  {Vasyuchka}, \citenamefont {Serga},\ and\ \citenamefont
  {Hillebrands}}]{ChumakNP15}%
  \BibitemOpen
  \bibfield  {author} {\bibinfo {author} {\bibfnamefont {A.}~\bibnamefont
  {Chumak}}, \bibinfo {author} {\bibfnamefont {V.}~\bibnamefont {Vasyuchka}},
  \bibinfo {author} {\bibfnamefont {A.}~\bibnamefont {Serga}}, \ and\ \bibinfo
  {author} {\bibfnamefont {B.}~\bibnamefont {Hillebrands}},\ }\href@noop {}
  {\bibfield  {journal} {\bibinfo  {journal} {Nature Physics}\ }\textbf
  {\bibinfo {volume} {11}},\ \bibinfo {pages} {453} (\bibinfo {year}
  {2015}{\natexlab{b}})}\BibitemShut {NoStop}%
\bibitem [{\citenamefont {Dzyaloshinsky}(1958)}]{DzyaloshinskyJPCS58}%
  \BibitemOpen
  \bibfield  {author} {\bibinfo {author} {\bibfnamefont {I.}~\bibnamefont
  {Dzyaloshinsky}},\ }\href@noop {} {\bibfield  {journal} {\bibinfo  {journal}
  {Journal of Physics and Chemistry of Solids}\ }\textbf {\bibinfo {volume}
  {4}},\ \bibinfo {pages} {241} (\bibinfo {year} {1958})}\BibitemShut {NoStop}%
\bibitem [{\citenamefont {Moriya}(1960)}]{MoriyaPR60}%
  \BibitemOpen
  \bibfield  {author} {\bibinfo {author} {\bibfnamefont {T.}~\bibnamefont
  {Moriya}},\ }\href@noop {} {\bibfield  {journal} {\bibinfo  {journal}
  {Physical Review}\ }\textbf {\bibinfo {volume} {120}},\ \bibinfo {pages} {91}
  (\bibinfo {year} {1960})}\BibitemShut {NoStop}%
\bibitem [{\citenamefont {Zakeri}\ \emph {et~al.}(2010)\citenamefont {Zakeri},
  \citenamefont {Zhang}, \citenamefont {Prokop}, \citenamefont {Chuang},
  \citenamefont {Sakr}, \citenamefont {Tang},\ and\ \citenamefont
  {Kirschner}}]{zakeri_asymmetric_2010}%
  \BibitemOpen
  \bibfield  {author} {\bibinfo {author} {\bibfnamefont {K.}~\bibnamefont
  {Zakeri}}, \bibinfo {author} {\bibfnamefont {Y.}~\bibnamefont {Zhang}},
  \bibinfo {author} {\bibfnamefont {J.}~\bibnamefont {Prokop}}, \bibinfo
  {author} {\bibfnamefont {T.-H.}\ \bibnamefont {Chuang}}, \bibinfo {author}
  {\bibfnamefont {N.}~\bibnamefont {Sakr}}, \bibinfo {author} {\bibfnamefont
  {W.~X.}\ \bibnamefont {Tang}}, \ and\ \bibinfo {author} {\bibfnamefont
  {J.}~\bibnamefont {Kirschner}},\ }\href {\doibase
  10.1103/PhysRevLett.104.137203} {\bibfield  {journal} {\bibinfo  {journal}
  {Physical Review Letters}\ }\textbf {\bibinfo {volume} {104}},\ \bibinfo
  {pages} {137203} (\bibinfo {year} {2010})}\BibitemShut {NoStop}%
\bibitem [{\citenamefont {Cort{\'e}s-Ortu{\~n}o}\ and\ \citenamefont
  {Landeros}(2013)}]{CorteOrtunoJPCM13}%
  \BibitemOpen
  \bibfield  {author} {\bibinfo {author} {\bibfnamefont {D.}~\bibnamefont
  {Cort{\'e}s-Ortu{\~n}o}}\ and\ \bibinfo {author} {\bibfnamefont
  {P.}~\bibnamefont {Landeros}},\ }\href
  {http://stacks.iop.org/0953-8984/25/i=15/a=156001} {\bibfield  {journal}
  {\bibinfo  {journal} {Journal of Physics: Condensed Matter}\ }\textbf
  {\bibinfo {volume} {25}},\ \bibinfo {pages} {156001} (\bibinfo {year}
  {2013})}\BibitemShut {NoStop}%
\bibitem [{\citenamefont {Ma}\ and\ \citenamefont {Zhou}(2014)}]{MaRSCA14}%
  \BibitemOpen
  \bibfield  {author} {\bibinfo {author} {\bibfnamefont {F.}~\bibnamefont
  {Ma}}\ and\ \bibinfo {author} {\bibfnamefont {Y.}~\bibnamefont {Zhou}},\
  }\href@noop {} {\bibfield  {journal} {\bibinfo  {journal} {RSC Adv.}\
  }\textbf {\bibinfo {volume} {4}},\ \bibinfo {pages} {46454} (\bibinfo {year}
  {2014})}\BibitemShut {NoStop}%
\bibitem [{\citenamefont {Di}\ \emph {et~al.}(2015)\citenamefont {Di},
  \citenamefont {Zhang}, \citenamefont {Lim}, \citenamefont {Ng}, \citenamefont
  {Kuok}, \citenamefont {Yu}, \citenamefont {Yoon}, \citenamefont {Qiu},\ and\
  \citenamefont {Yang}}]{DiPRL15}%
  \BibitemOpen
  \bibfield  {author} {\bibinfo {author} {\bibfnamefont {K.}~\bibnamefont
  {Di}}, \bibinfo {author} {\bibfnamefont {V.~L.}\ \bibnamefont {Zhang}},
  \bibinfo {author} {\bibfnamefont {H.~S.}\ \bibnamefont {Lim}}, \bibinfo
  {author} {\bibfnamefont {S.~C.}\ \bibnamefont {Ng}}, \bibinfo {author}
  {\bibfnamefont {M.~H.}\ \bibnamefont {Kuok}}, \bibinfo {author}
  {\bibfnamefont {J.}~\bibnamefont {Yu}}, \bibinfo {author} {\bibfnamefont
  {J.}~\bibnamefont {Yoon}}, \bibinfo {author} {\bibfnamefont {X.}~\bibnamefont
  {Qiu}}, \ and\ \bibinfo {author} {\bibfnamefont {H.}~\bibnamefont {Yang}},\
  }\href@noop {} {\bibfield  {journal} {\bibinfo  {journal} {Physical review
  letters}\ }\textbf {\bibinfo {volume} {114}},\ \bibinfo {pages} {047201}
  (\bibinfo {year} {2015})}\BibitemShut {NoStop}%
\bibitem [{\citenamefont {Iguchi}\ \emph {et~al.}(2015)\citenamefont {Iguchi},
  \citenamefont {Uemura}, \citenamefont {Ueno},\ and\ \citenamefont
  {Onose}}]{IguchiPRB15}%
  \BibitemOpen
  \bibfield  {author} {\bibinfo {author} {\bibfnamefont {Y.}~\bibnamefont
  {Iguchi}}, \bibinfo {author} {\bibfnamefont {S.}~\bibnamefont {Uemura}},
  \bibinfo {author} {\bibfnamefont {K.}~\bibnamefont {Ueno}}, \ and\ \bibinfo
  {author} {\bibfnamefont {Y.}~\bibnamefont {Onose}},\ }\href {\doibase
  10.1103/PhysRevB.92.184419} {\bibfield  {journal} {\bibinfo  {journal} {Phys.
  Rev. B}\ }\textbf {\bibinfo {volume} {92}},\ \bibinfo {pages} {184419}
  (\bibinfo {year} {2015})}\BibitemShut {NoStop}%
\bibitem [{\citenamefont {Hertel}(2013)}]{hertel_curvature-induced_2013}%
  \BibitemOpen
  \bibfield  {author} {\bibinfo {author} {\bibfnamefont {R.}~\bibnamefont
  {Hertel}},\ }\href {\doibase 10.1142/S2010324713400092} {\bibfield  {journal}
  {\bibinfo  {journal} {SPIN}\ }\textbf {\bibinfo {volume} {3}},\ \bibinfo
  {pages} {1340009} (\bibinfo {year} {2013})}\BibitemShut {NoStop}%
\bibitem [{\citenamefont {Gaididei}\ \emph {et~al.}(2017)\citenamefont
  {Gaididei}, \citenamefont {Goussev}, \citenamefont {Kravchuk}, \citenamefont
  {Pylypovskyi}, \citenamefont {Robbins}, \citenamefont {Sheka}, \citenamefont
  {Slastikov},\ and\ \citenamefont {Vasylkevych}}]{GaidideiArxiv17}%
  \BibitemOpen
  \bibfield  {author} {\bibinfo {author} {\bibfnamefont {Y.}~\bibnamefont
  {Gaididei}}, \bibinfo {author} {\bibfnamefont {A.}~\bibnamefont {Goussev}},
  \bibinfo {author} {\bibfnamefont {V.~P.}\ \bibnamefont {Kravchuk}}, \bibinfo
  {author} {\bibfnamefont {O.~V.}\ \bibnamefont {Pylypovskyi}}, \bibinfo
  {author} {\bibfnamefont {J.}~\bibnamefont {Robbins}}, \bibinfo {author}
  {\bibfnamefont {D.~D.}\ \bibnamefont {Sheka}}, \bibinfo {author}
  {\bibfnamefont {V.}~\bibnamefont {Slastikov}}, \ and\ \bibinfo {author}
  {\bibfnamefont {S.}~\bibnamefont {Vasylkevych}},\ }\href@noop {} {\bibfield
  {journal} {\bibinfo  {journal} {arXiv preprint arXiv:1701.01691}\ } (\bibinfo
  {year} {2017})}\BibitemShut {NoStop}%
\bibitem [{\citenamefont {Sheka}\ \emph {et~al.}(2015)\citenamefont {Sheka},
  \citenamefont {Kravchuk}, \citenamefont {Yershov},\ and\ \citenamefont
  {Gaididei}}]{ShekaPRB17}%
  \BibitemOpen
  \bibfield  {author} {\bibinfo {author} {\bibfnamefont {D.~D.}\ \bibnamefont
  {Sheka}}, \bibinfo {author} {\bibfnamefont {V.~P.}\ \bibnamefont {Kravchuk}},
  \bibinfo {author} {\bibfnamefont {K.~V.}\ \bibnamefont {Yershov}}, \ and\
  \bibinfo {author} {\bibfnamefont {Y.}~\bibnamefont {Gaididei}},\ }\href
  {\doibase 10.1103/PhysRevB.92.054417} {\bibfield  {journal} {\bibinfo
  {journal} {Phys. Rev. B}\ }\textbf {\bibinfo {volume} {92}},\ \bibinfo
  {pages} {054417} (\bibinfo {year} {2015})}\BibitemShut {NoStop}%
\bibitem [{\citenamefont {Ot{\'a}lora}\ \emph {et~al.}(2016)\citenamefont
  {Ot{\'a}lora}, \citenamefont {Yan}, \citenamefont {Schultheiss},
  \citenamefont {Hertel},\ and\ \citenamefont {K{\'a}kay}}]{OtaloraPRL16}%
  \BibitemOpen
  \bibfield  {author} {\bibinfo {author} {\bibfnamefont {J.~A.}\ \bibnamefont
  {Ot{\'a}lora}}, \bibinfo {author} {\bibfnamefont {M.}~\bibnamefont {Yan}},
  \bibinfo {author} {\bibfnamefont {H.}~\bibnamefont {Schultheiss}}, \bibinfo
  {author} {\bibfnamefont {R.}~\bibnamefont {Hertel}}, \ and\ \bibinfo {author}
  {\bibfnamefont {A.}~\bibnamefont {K{\'a}kay}},\ }\href@noop {} {\bibfield
  {journal} {\bibinfo  {journal} {Physical review letters}\ }\textbf {\bibinfo
  {volume} {117}},\ \bibinfo {pages} {227203} (\bibinfo {year}
  {2016})}\BibitemShut {NoStop}%
\bibitem [{\citenamefont {Ot\'alora}\ \emph {et~al.}(2017)\citenamefont
  {Ot\'alora}, \citenamefont {Yan}, \citenamefont {Schultheiss}, \citenamefont
  {Hertel},\ and\ \citenamefont {K\'akay}}]{OtaloraPRB17}%
  \BibitemOpen
  \bibfield  {author} {\bibinfo {author} {\bibfnamefont {J.~A.}\ \bibnamefont
  {Ot\'alora}}, \bibinfo {author} {\bibfnamefont {M.}~\bibnamefont {Yan}},
  \bibinfo {author} {\bibfnamefont {H.}~\bibnamefont {Schultheiss}}, \bibinfo
  {author} {\bibfnamefont {R.}~\bibnamefont {Hertel}}, \ and\ \bibinfo {author}
  {\bibfnamefont {A.}~\bibnamefont {K\'akay}},\ }\href {\doibase
  10.1103/PhysRevB.95.184415} {\bibfield  {journal} {\bibinfo  {journal} {Phys.
  Rev. B}\ }\textbf {\bibinfo {volume} {95}},\ \bibinfo {pages} {184415}
  (\bibinfo {year} {2017})}\BibitemShut {NoStop}%
\bibitem [{\citenamefont {Polder}(1949)}]{PolderMJS49}%
  \BibitemOpen
  \bibfield  {author} {\bibinfo {author} {\bibfnamefont {D.}~\bibnamefont
  {Polder}},\ }\href@noop {} {\bibfield  {journal} {\bibinfo  {journal} {The
  London, Edinburgh, and Dublin Philosophical Magazine and Journal of Science}\
  }\textbf {\bibinfo {volume} {40}},\ \bibinfo {pages} {99} (\bibinfo {year}
  {1949})}\BibitemShut {NoStop}%
\bibitem [{\citenamefont {Hurben}\ and\ \citenamefont
  {Patton}(1998)}]{HurbenJAP98}%
  \BibitemOpen
  \bibfield  {author} {\bibinfo {author} {\bibfnamefont {M.}~\bibnamefont
  {Hurben}}\ and\ \bibinfo {author} {\bibfnamefont {C.}~\bibnamefont
  {Patton}},\ }\href@noop {} {\bibfield  {journal} {\bibinfo  {journal}
  {Journal of Applied Physics}\ }\textbf {\bibinfo {volume} {83}},\ \bibinfo
  {pages} {4344} (\bibinfo {year} {1998})}\BibitemShut {NoStop}%
\bibitem [{\citenamefont {Arias}\ and\ \citenamefont
  {Mills}(1999)}]{AriasPRB99}%
  \BibitemOpen
  \bibfield  {author} {\bibinfo {author} {\bibfnamefont {R.}~\bibnamefont
  {Arias}}\ and\ \bibinfo {author} {\bibfnamefont {D.~L.}\ \bibnamefont
  {Mills}},\ }\href {\doibase 10.1103/PhysRevB.60.7395} {\bibfield  {journal}
  {\bibinfo  {journal} {Phys. Rev. B}\ }\textbf {\bibinfo {volume} {60}},\
  \bibinfo {pages} {7395} (\bibinfo {year} {1999})}\BibitemShut {NoStop}%
\bibitem [{\citenamefont {Landeros}\ \emph {et~al.}(2008)\citenamefont
  {Landeros}, \citenamefont {Arias},\ and\ \citenamefont
  {Mills}}]{LanderosPRB08}%
  \BibitemOpen
  \bibfield  {author} {\bibinfo {author} {\bibfnamefont {P.}~\bibnamefont
  {Landeros}}, \bibinfo {author} {\bibfnamefont {R.~E.}\ \bibnamefont {Arias}},
  \ and\ \bibinfo {author} {\bibfnamefont {D.}~\bibnamefont {Mills}},\
  }\href@noop {} {\bibfield  {journal} {\bibinfo  {journal} {Physical Review
  B}\ }\textbf {\bibinfo {volume} {77}},\ \bibinfo {pages} {214405} (\bibinfo
  {year} {2008})}\BibitemShut {NoStop}%
\bibitem [{\citenamefont {Gallardo}\ \emph {et~al.}(2014)\citenamefont
  {Gallardo}, \citenamefont {Banholzer}, \citenamefont {Wagner}, \citenamefont
  {K{\"o}rner}, \citenamefont {Lenz}, \citenamefont {Farle}, \citenamefont
  {Lindner}, \citenamefont {Fassbender},\ and\ \citenamefont
  {Landeros}}]{GallardoNJP14}%
  \BibitemOpen
  \bibfield  {author} {\bibinfo {author} {\bibfnamefont {R.}~\bibnamefont
  {Gallardo}}, \bibinfo {author} {\bibfnamefont {A.}~\bibnamefont {Banholzer}},
  \bibinfo {author} {\bibfnamefont {K.}~\bibnamefont {Wagner}}, \bibinfo
  {author} {\bibfnamefont {M.}~\bibnamefont {K{\"o}rner}}, \bibinfo {author}
  {\bibfnamefont {K.}~\bibnamefont {Lenz}}, \bibinfo {author} {\bibfnamefont
  {M.}~\bibnamefont {Farle}}, \bibinfo {author} {\bibfnamefont
  {J.}~\bibnamefont {Lindner}}, \bibinfo {author} {\bibfnamefont
  {J.}~\bibnamefont {Fassbender}}, \ and\ \bibinfo {author} {\bibfnamefont
  {P.}~\bibnamefont {Landeros}},\ }\href@noop {} {\bibfield  {journal}
  {\bibinfo  {journal} {New Journal of Physics}\ }\textbf {\bibinfo {volume}
  {16}},\ \bibinfo {pages} {023015} (\bibinfo {year} {2014})}\BibitemShut
  {NoStop}%
\bibitem [{\citenamefont {Twisselmann}\ and\ \citenamefont
  {McMichael}(2003)}]{TwisselmannJAP03}%
  \BibitemOpen
  \bibfield  {author} {\bibinfo {author} {\bibfnamefont {D.}~\bibnamefont
  {Twisselmann}}\ and\ \bibinfo {author} {\bibfnamefont {R.~D.}\ \bibnamefont
  {McMichael}},\ }\href@noop {} {\bibfield  {journal} {\bibinfo  {journal}
  {Journal of Applied Physics}\ }\textbf {\bibinfo {volume} {93}},\ \bibinfo
  {pages} {6903} (\bibinfo {year} {2003})}\BibitemShut {NoStop}%
\bibitem [{\citenamefont {Kalarickal}\ \emph {et~al.}(2006)\citenamefont
  {Kalarickal}, \citenamefont {Krivosik}, \citenamefont {Wu}, \citenamefont
  {Patton}, \citenamefont {Schneider}, \citenamefont {Kabos}, \citenamefont
  {Silva},\ and\ \citenamefont {Nibarger}}]{KalarickalJAP06}%
  \BibitemOpen
  \bibfield  {author} {\bibinfo {author} {\bibfnamefont {S.~S.}\ \bibnamefont
  {Kalarickal}}, \bibinfo {author} {\bibfnamefont {P.}~\bibnamefont
  {Krivosik}}, \bibinfo {author} {\bibfnamefont {M.}~\bibnamefont {Wu}},
  \bibinfo {author} {\bibfnamefont {C.~E.}\ \bibnamefont {Patton}}, \bibinfo
  {author} {\bibfnamefont {M.~L.}\ \bibnamefont {Schneider}}, \bibinfo {author}
  {\bibfnamefont {P.}~\bibnamefont {Kabos}}, \bibinfo {author} {\bibfnamefont
  {T.~J.}\ \bibnamefont {Silva}}, \ and\ \bibinfo {author} {\bibfnamefont
  {J.~P.}\ \bibnamefont {Nibarger}},\ }\href@noop {} {\bibfield  {journal}
  {\bibinfo  {journal} {Journal of Applied Physics}\ }\textbf {\bibinfo
  {volume} {99}},\ \bibinfo {pages} {093909} (\bibinfo {year}
  {2006})}\BibitemShut {NoStop}%
\bibitem [{Note1()}]{Note1}%
  \BibitemOpen
  \bibinfo {note} {Note that the same stiffness fields appear in the dispersion
  relation ($\omega _{2D}[\protect \mathaccentV {vec}17E{k}]=\omega
  _{M_s}\protect \sqrt {h_X[\protect \mathaccentV {vec}17E{k}] h_Y[\protect
  \mathaccentV {vec}17E{k}]}$).}\BibitemShut {Stop}%
\bibitem [{\citenamefont {Kittel}(1986)}]{Kittel:ISSP}%
  \BibitemOpen
  \bibfield  {author} {\bibinfo {author} {\bibfnamefont {C.}~\bibnamefont
  {Kittel}},\ }\href@noop {} {\emph {\bibinfo {title} {{Introduction to Solid
  State Physics}}}},\ \bibinfo {edition} {6th}\ ed.\ (\bibinfo  {publisher}
  {John Wiley \& Sons, Inc.},\ \bibinfo {address} {New York},\ \bibinfo {year}
  {1986})\BibitemShut {NoStop}%
\end{thebibliography}
\bibliographystyle{apsrev4-1}

\end{document}